\documentclass[fleqn,usenatbib,useAMS]{mnras}

\usepackage{newtxtext,newtxmath}

\usepackage[T1]{fontenc}

\DeclareRobustCommand{\VAN}[3]{#2}
\let\VANthebibliography\thebibliography
\def\thebibliography{\DeclareRobustCommand{\VAN}[3]{##3}\VANthebibliography}

\usepackage{hyperref}
\usepackage{subcaption}
\usepackage{graphicx}	
\usepackage{amsmath}	
\usepackage{tablefootnote}
\usepackage{longtable}
\usepackage{orcidlink}

\title[RIDDLER: SNe~Ia standardisation]{Quantitative modelling of type Ia supernovae spectral time series III:\\Implications for type Ia supernovae standardisation in cosmology}

\author[M. R. Magee]{
M. R. Magee$^{1}$\orcidlink{0000-0002-0629-8931}\thanks{E-mail: mrmagee.astro@gmail.com}
\\
$^{1}$Department of Physics, University of Warwick, Gibbet Hill Road, Coventry CV4 7AL, UK \\ 
}

\date{Accepted XXX. Received YYY; in original form ZZZ}

\pubyear{2026}

\begin{document}
\label{firstpage}
\pagerange{\pageref{firstpage}--\pageref{lastpage}}
\maketitle

\begin{abstract}
The physics driving type Ia supernovae (SNe~Ia) standardisation in cosmology remains poorly-understood. Recent advances however mean that it is now possible to systematically analyse the explosion properties of large numbers of cosmological SNe~Ia. To that end we use \textsc{riddler}, a machine learning based framework for rapidly modelling SNe~Ia based on realistic explosion simulations, to perform quantitative spectral modelling of the Zwicky Transient Facility SN~Ia DR2 sample and determine their best-fitting explosion mechanism(s). We find that approximately two thirds of our sample is best reproduced by sub-Chandrasekhar mass explosions. Analysing their light curve and host galaxy properties, we find that Chandrasekhar mass explosions are not favoured for the fastest-evolving SNe~Ia, while sub-Chandrasekhar mass explosions are favoured for the reddest SNe~Ia. Due to the differences in their environments, selecting SNe~Ia in massive, passive galaxies could produce a homogeneous sample of violent merger SNe~Ia. We show that standardising each explosion mechanism independently reduces scatter in distance estimates and previously claimed environmental and non-linear light curve shape corrections may be due to changes in the relative populations of different explosion mechanisms. Although a step forward towards understanding SNe~Ia physics in cosmology, we highlight a number of limitations affecting our conclusions, including sample biases and small numbers. We therefore cannot assess the statistical significance of our results and they should be treated with caution. Larger and more uniformly observed samples will be key to determining the significance of any trends hinted at here. 
\end{abstract}




\section{Introduction}
\label{sect:intro}

Towards the end of the 20$^{\rm{th}}$ century, the standardisable nature of type Ia supernovae (SNe~Ia) enabled the discovery of the accelerating expansion of the Universe \citep{riess--expansion, perlmutter99}. Subsequent cosmology surveys have increased sample sizes and reduced systematic uncertainties to yield more precise measurements of the cosmic expansion rate, $H_0$ (e.g. \citealt{jones--19, brout--22, des--24}). The most recent release from the SH0ES team reports a value of $H_0 = 73.04\pm1.04$~km~s$^{-1}~\rm{Mpc}^{-1}$ \citep{riess--22}, which is higher than that measured from the Planck cosmic microwave background and in tension at a 5$\sigma$ level \citep{planck--20}. The source of this discrepancy may lie in unknown physics lacking from cosmological models or unknown systematics affecting SNe~Ia standardisation and/or sample selection. 

\par

In an effort to resolve this discrepancy, the goal of modern SN cosmology studies is typically to reduce scatter in so-called Hubble residuals -- the difference between observed distance modulii measured from SNe~Ia light curves and those predicted by cosmological models. Uncorrected, the peak absolute magnitudes of SNe~Ia show a scatter of $\sim$0.35~mag, but following standardisation this can be reduced to $\sim$0.15~mag \citep{tripp--98}. Standardisation of SNe~Ia exploits the observed correlations between the peak magnitude, light curve width, and colour. Brighter SNe~Ia are both slower evolving and show bluer colours \citep{ia--decline, phillips--93, tripp--98}. Further correlations are also observed between light curve properties of SNe~Ia and the host galaxy environment, leading to additional corrections being applied. The most common correction is the mass step, SNe~Ia in more massive galaxies are typically brighter than those in less massive galaxies after standardisation \citep{kelly--10, sullivan--10, gupta--11, childress--13}, but corrections have also been suggested due to the galaxy star formation rate \citep{sullivan--06, rigault--13, rigault--20}, colour \citep{childress--13, roman--18, kelsey--21, kelsey--23, ginolin--25a, ginolin--25b}, morphology \citep{hamuy--00, pruzhinskaya--20}, and metallicity \citep{gallager--05, gallagher--08, dandrea--11, pan--14}. The significance of environmental corrections vary depending on exactly which tracer is used \citep{briday--22}.

\par

Although correlations between SN properties and host galaxy properties are now well established, the origins of these relationships are unclear and lingering scatter in Hubble residuals still remains. \cite{brout--21} argue that differences in dust laws among SNe~Ia host galaxies are the dominant component driving correlations with the host galaxy properties. Conversely, \cite{uddin--20, uddin--24} use observations across the optical and near-infrared to argue that dust does not play a significant role. Other studies of SNe~Ia light curves in the near-infrared have also debated the impact of dust \citep{johansson--21, thorp--21, grayling--24}. Alternatively, many studies assume that these relationships reflect intrinsic differences among SNe~Ia populations that are related to the progenitors and their prevalence in different environments. Commonly this takes the form of a two component model, whereby SNe~Ia arise from both young and old progenitors \citep{mannucci--06, smith--12, maoz--14, nicholas--21, briday--22, wiseman--22}. 

\par

While it is widely accepted that SNe~Ia result from thermonuclear explosions of white dwarfs in binary systems, multiple progenitor systems and explosion mechanisms have been proposed (see \citealt{ruiter--25} for a recent review). If multiple channels, and indeed more than two, are responsible for producing the SNe~Ia found in cosmological samples, they may not share the same global standardisation parameters and hence additional systematic scatter could be introduced \citep{wojtak--22}. \cite{gonzalez-gaitan--21} investigate the effects of varying the luminosity-colour relations among SNe~Ia and argue for at least two distinct populations. Similarly, \cite{ginolin--25b} find that the luminosity-stretch (light curve shape) relation is non-linear, as opposed to the linear relationship often assumed, and shows two distinct modes that may be reflective of different populations. \cite{ramaiya--25} categorise their sample of SNe~Ia into three separate populations, depending on the host galaxy star formation rate and mass, and also find evidence of significant variation among their colour-luminosity relations. The presence of multiple populations has strong implications for precision cosmology as correctly accounting for different underlying populations can lead to a significant reduction in the disagreement between $H_0$ measurements from SNe~Ia and the cosmic microwave background \citep{martins--25, wojtak--25}.

\par

Rather than implementing multiple or complex standardisation corrections for different SNe~Ia populations, selecting sub-samples that share a common origin could simplify this process. Some studies have argued that blue SNe~Ia are more homogeneous than the general population and therefore could provide a more robust sample for future surveys \citep{kelsey--21, gall--24}. These SNe~Ia should be the least affected by dust extinction hence any variations in their colours may be intrinsic \citep{mandel--17, mandel--22}. Further attempts have also been made to link sub-samples to intrinsic properties of the SN. By generating composite spectra of SNe~Ia, \cite{siebery--20a} show that those with higher ejecta velocities are typically brighter following correction than those with lower velocities. Conversely, other studies have found no significant difference between the Hubble residuals of normal and high velocity SNe~Ia \citep{dettman--21, pan--24}. \cite{burgaz--26} find similar results, but also argue that environmental corrections differ significantly between the two velocity samples. This could suggest that the explosion properties of SNe~Ia are closely related to their local environments and therefore SNe~Ia in cosmology samples should be selected based on their explosion physics. At the same time, \cite{sarin--26} use a physically-motivated model to fit the light curves of 2\,205 SNe~Ia and investigate the explosion properties, including $^{56}$Ni and ejecta masses. \cite{sarin--26} argue that this model provides a physical basis for standardisation that may remove the need for environmental corrections and that a single dominant explosion mechanism operates across a range of progenitor masses.

\par

Determining the explosion physics of SNe~Ia is non-trivial, and it is currently unclear whether proposed sub-samples truly represent distinct populations. Many studies have shown that different explosion mechanisms can produce similar observables, making distinguishing between them for individual SNe~Ia a challenging prospect \citep{hoeflich--96a, sim--13, dessart--14a, blondin--17, hoeflich--2017, goldstein--18, shen--18, shen--21, blondin--22}. Detailed empirical spectroscopic modelling of individual SNe~Ia can also reveal useful insights into the explosion physics, but this process is time and computationally intensive, and difficult to apply robustly to larger samples \citep{stehle--05, mazzali--07, mazzali--08, tanaka--11, mazzali--14, sasdelli--14, ashall--16, heringer--17, 12bwh, heringer--19, harvey--23, yadavalli--24}. To overcome these limitations, \cite{magee--24} present \textsc{riddler}, a machine learning-based method for automatic and quantitative fitting of SNe~Ia spectral time series. The training data for \textsc{riddler} is directly based on predictions from realistic explosion simulations and was recently updated to include a wider range of explosion mechanisms \citep{magee--26a}. With \textsc{riddler} it is therefore possible to quantitatively assess the relative likelihood of different explosion mechanisms and hence the existence of multiple populations among cosmology samples of SNe~Ia, which may require specific standardisation corrections.

\par

In this paper, we present the first attempt at systematically modelling the spectra of a large number of cosmological SNe~Ia in a consistent manner to extract their explosion physics. We use the recently updated \textsc{riddler} to perform fits to the Zwicky Transient Facility (ZTF) SN~Ia DR2 sample \citep{rigault--25} and investigate correlations between properties of the explosion and cosmology standardisation parameters. We also highlight a number of limitations and considerations that should be taken into account when applying such an approach to other cosmology samples. In Sect.~\ref{sect:sample} we discuss the samples used in this work. Section~\ref{sect:riddler} briefly describes \textsc{riddler} and the process of fitting SNe~Ia spectral time series. In Sect.~\ref{sect:results} we present the results of our \textsc{riddler} fits. Section~\ref{sect:discussion} discusses our results, including biases in our samples and the impact on SNe~Ia standardisation. Finally, we conclude in Sect.~\ref{sect:conclusions}.

%

\section{Zwicky Transient Facility sample}
\label{sect:sample}

\begin{table*}
\begin{center}
\caption{Summary of the cuts applied to the ZTF SN~Ia DR2 sample.}
\label{tab:cuts}
\begin{tabular}{lrrrr}
\hline
\textbf{Condition} & \textbf{Objects} & \textbf{Percentage} & \textbf{Spectra} & \textbf{Percentage}  \\
& \textbf{remaining} & \textbf{removed} & \textbf{remaining} & \textbf{removed} \\
\hline
\hline
Starting sample                                               & 3\,628 & $\cdots$ & 5\,138 & $\cdots$           \\
Cosmology cuts                                                & 2652   & 27\% & 3698 & 28\% \\
$z \leq 0.06$                                                 & 979    & 63\% & 1557 & 58\% \\
$-11$\,d $\leq$ phase $\leq$ $+5$\,d                          & 787    & 20\% & 1098 & 29\% \\
SNR $\geq$10                                                  & 517    & 34\% & 652  & 41\% \\
Resolution $\geq$300                                          & 244    & 53\% & 275  & 58\% \\
At least 2 spectra                                            & 28 & 89\% & 59 & 79\% \\
\hline
\hline
\end{tabular}
\end{center}
\end{table*}

The ZTF SN~Ia DR2 includes all spectroscopically confirmed SNe~Ia observed by ZTF between March 2018 and December 2020 \citep{rigault--25}. The light curve, spectroscopic, and host galaxy properties have been studied extensively (see \citealt{rigault--25} and references therein) and, along with the data, are all publicly available. This makes the ZTF SN~Ia DR2 an excellent test case to examine the process of quantitative spectral modelling of cosmological SNe~Ia and exploring differences in their explosion properties. Table~\ref{tab:cuts} outlines the cuts (detailed below) applied to produce the samples of SNe~Ia spectra that we fit with \textsc{riddler}.

\par

The data release provided by \cite{rigault--25} includes 5\,138 spectra of 3\,628 SNe~Ia. \cite{rigault--25} outline a series of cosmology selection criteria based on the overall quality of the SN light curve and SALT2 \citep{guy--07} fit parameters. Applying these cuts reduces our sample to 3\,698 spectra of 2\,652 SNe~Ia. We set a redshift limit of $z \leq 0.06$, as this is the redshift at which ZTF is expected to be spectroscopically complete for normal SNe~Ia \citep{amenouche--25}, bringing our sample to 1\,557 spectra of 979 SNe~Ia. 

\par

We apply a number of additional cuts based on the available spectra. The training data for \textsc{riddler} covers 5 -- 30\,d after explosion and typically assumes rise times of $\sim$16 -- 25\,d. We therefore select only those spectra with phases (relative to maximum light) in the range of $-11$ -- $+5$\,d. This ensures that all observed spectra fall within the correct phase range of our training data regardless of the explosion epoch being fit. Following this cut, our sample is reduced to 1\,098 spectra of 787 SNe~Ia. \cite{magee--26a} show that spectra with low signal-to-noise ratios (SNRs) may produce unreliable fits. We estimate the SNR of each spectrum by taking the median and standard deviation of the flux across 10 wavelength bins and average this over the entire spectrum. Removing spectra with average SNRs $\textless$10 reduces our sample to 652 spectra of 517 SNe~Ia. \cite{magee--26a} also show that low resolution spectra may produce similarly unreliable fits. Approximately 60~per~cent of all spectra released as part of the ZTF SN~Ia DR2 were observed with the low resolution SED machine \citep{sedm}, which was designed primarily for spectroscopic classification and observing efficiency rather than detailed spectroscopic analysis. Removing spectra with resolutions $\textless$300 (at 6\,000~\AA) therefore represents a significant cut to our sample, giving only 275 spectra of 244 SNe~Ia. 

\par

Our final cut is based on the number of spectra available for each SN. The time evolution of spectral features provides tighter constraints on the ejecta model and explosion properties of the SN compared to fitting a single spectrum alone. While it may be possible to fit a single spectrum and find a good match, without additional data it is unclear whether the model would produce a similarly good match at other (unseen) epochs. Therefore fits to a single spectrum may not be robust -- adding more spectra could alter which explosion scenario is found to produce the best agreement overall. Even fitting two spectra may not provide robust results for a given SN as ideally observations would cover three important phases during the spectral evolution: pre-maximum, maximum, and post-maximum. Nevertheless, fits to a limited number of spectra still provide insight into the models that could plausibly reproduce that SN or that could be excluded. For cosmology surveys, typically only a single spectrum is required to classify a new transient as a SN~Ia and therefore a requirement of multiple spectra would severely limit our sample. Indeed, only three SNe~Ia have three spectra that meet our selection criteria and none have more than three spectra. We therefore split our SNe into primary and secondary samples. Our primary sample consists of those SNe~Ia with at least two spectra. This gives 59 spectra of 28 SNe~Ia. Our secondary sample includes the remaining set of 216 SNe~Ia and 216 spectra meeting the previous selection criteria. We analyse both samples and present our results accordingly, but caution over the robustness of the fits, particularly for the secondary sample.

\par

With our samples selected, fitting with \textsc{riddler} requires that our spectra are calibrated in physical flux units. For each observed spectrum, we generate a template spectrum based on the SALT2 parameters and fit a low order polynomial to estimate the continuum. Observed spectra are then scaled to match this continuum shape. This ensures that both the absolute and relative flux calibration is consistent with the observed photometry. We note that \cite{ganot--26} present a sample of flux calibrated spectra from the ZTF SN~Ia DR2 however this sample is limited to only SED machine spectra and therefore is not used here.

\par

\begin{figure*}
\centering
\includegraphics[width=\textwidth]{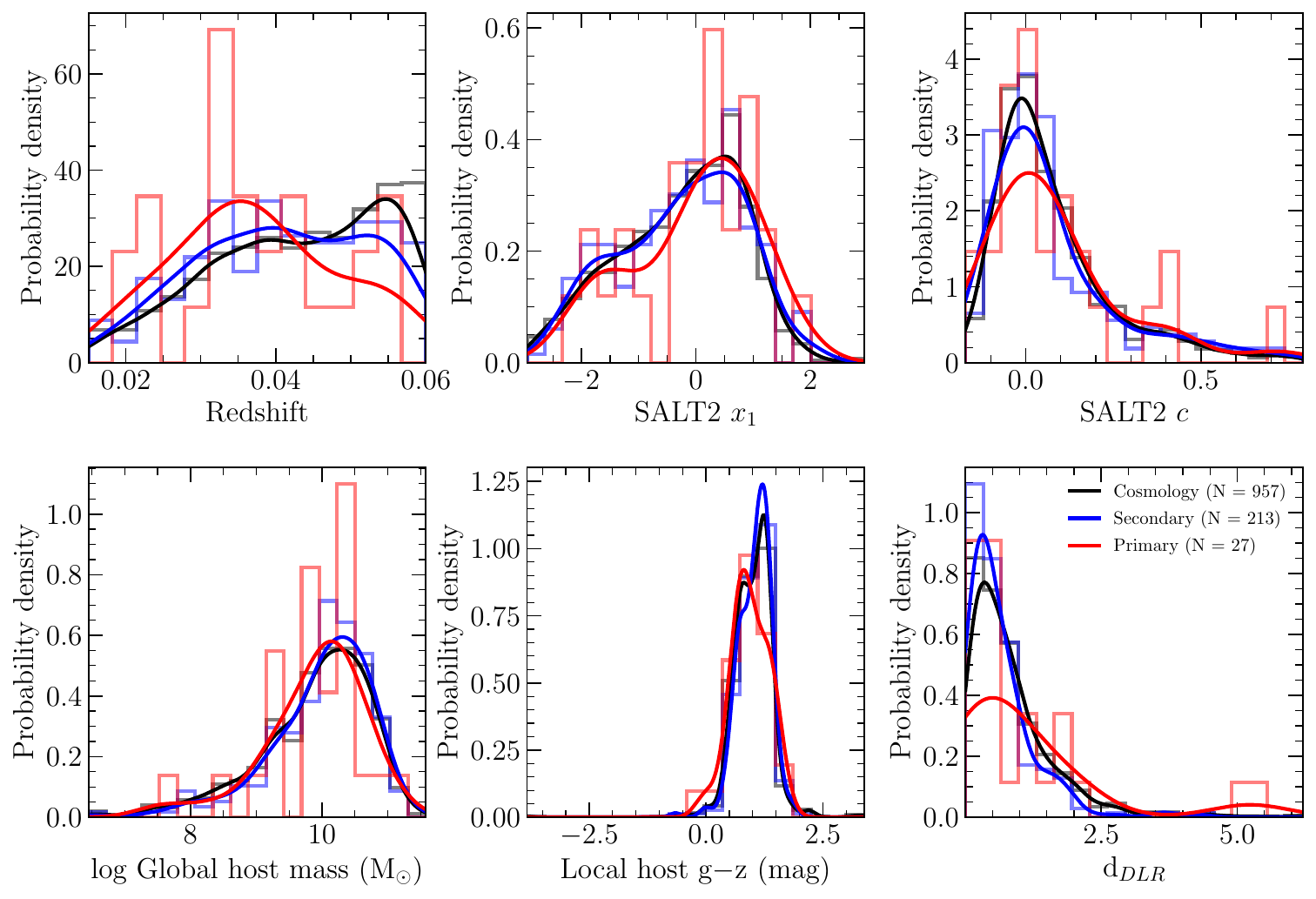}
\caption{Distributions of SN parameters for our primary (red) and secondary (blue) samples, and the parent ZTF SN~Ia DR2 cosmology sample (black). Faded lines show histograms of each parameter and solid lines show smoothed kernel density estimates. Note that some SNe~Ia have been excluded as they do not have measured host galaxy properties.
}
\label{fig:sample_distributions}
\centering
\end{figure*}

In Fig.~\ref{fig:sample_distributions} we show the distributions of various parameters for the initial 979 cosmological SNe~Ia (note that some SNe are not included here as they do not have measured host galaxy properties), and our primary and secondary samples. All parameter values are taken from the ZTF SN~Ia DR2 \citep{rigault--25}. Unsurprisingly, given the requirement of high SNR spectra, our primary and secondary samples show a slight bias towards lower redshifts than the parent cosmology sample, but this is not statistically significant. All other parameters show similar distributions between the three samples, indicating that our primary and secondary samples span the full range of values found in the parent cosmology sample.

\par

We note that despite the similarities in parameter distributions highlighted by Fig.~\ref{fig:sample_distributions}, our sample is inherently biased. Most notably, our primary sample includes only those SNe~Ia with at least two spectra. While ZTF is spectroscopically complete up to a redshift of $z = 0.06$ \citep{amenouche--25}, clearly not all SNe have been observed spectroscopically multiple times. The selection of targets for additional spectroscopy beyond classification is arbitrary and may be based on specific light curve, spectral, or host galaxy features. Even for our secondary sample, we require high resolution spectra and the selection of targets for high resolution spectra, as opposed to SED machine or other instruments, may be similarly arbitrary. Finally, we also require high SNR spectra and this will bias our sample towards intrinsically brighter and/or more nearby SNe~Ia. For these reasons, and the previously mentioned limitations associated with fitting typically one or two spectra per SN, we are unable to robustly assess the statistical significance of any trends or correlations found between observable properties and the types of explosion mechanisms found by our fits. Nevertheless, our samples span similar ranges to the full cosmology sample and therefore are useful for exploring the ranges of properties that could plausibly be found among different types of explosion mechanisms.

%

\section{riddler}
\label{sect:riddler}

To fit the observations described in Sect.~\ref{sect:sample} we use \textsc{riddler}, which was first presented by \cite{magee--24} and subsequently updated by \cite{magee--26a}. Here we provide a brief overview and refer the reader to those works for more detailed descriptions.

\par

\textsc{riddler} is a machine learning-based framework designed for simultaneously fitting spectral time series of SNe~Ia. \textsc{tardis} \citep{tardis} is used to generate the synthetic spectra that form the basis of the \textsc{riddler} machine learning training data. \textsc{tardis} requires a number of input parameters for each simulation, including the density and composition of the ejecta, the time since explosion ($t_{\mathrm{exp}})$, the luminosity, and the inner boundary velocity ($v_i$). To construct our \textsc{tardis} models we define a set of 16 input parameters as:
\begin{itemize}
    \item Explosion strength, $ES$
    \item Total ejecta mass, $M_{\mathrm{ej}}$,
    \item Fraction of the ejecta contained within the core, $f_C$,
    \item Fraction of $f_C$ burned to heavy elements, $f_{C}^{b}$,
    \item Fraction of $f_{C}^{b}$ burned to iron-group elements, $f_{C}^{\mathrm{IGE}}$,
    \item Fraction of $f_{C}^{\mathrm{IGE}}$ burned to $^{56}$Ni, $f_{C}^{\textrm{Ni}}$,
    \item Fraction of the shell, $f_S$, burned to heavy elements, $f_{S}^{b}$,
    \item Fraction of $f_{S}^{b}$ burned to iron-group elements, $f_{S}^{\mathrm{IGE}}$,
    \item Fraction of $f_{S}^{\mathrm{IGE}}$ burned to $^{56}$Ni, $f_{S}^{\textrm{Ni}}$,
    \item Fraction of carbon/oxygen in the unburned shell material, $f_{S}^{C/O}$,
    \item Kinetic energy, $KE$,
    \item Rise time, $t_r$,
    \item Luminosity at maximum light, $L_{\mathrm{max}}$,
    \item Velocity at maximum light, $v_{\mathrm{max}}$,
    \item Velocity gradient, $\Delta v$,
    \item Time since explosion, $t_{\mathrm{exp}}$.
\end{itemize}
The first 11 parameters are used to determine the structure and composition of the ejecta, which contains a `core' and a `shell'. Following predictions from explosion simulations, including pure deflagrations (DEF; \citealt{fink-2014}), delayed detonations (DDT; \citealt{seitenzahl--13}), double detonations (DOD; \citealt{gronow--20}), gravitationally confined detonations (GCD; \citealt{lach--22b}), and violent mergers (VM; \citealt{pakmor--10, pakmor-2012, kromer--13b}), we randomly sample each parameter. The explosion strength parameter is sampled independently (within the boundaries set by the explosion simulations) while most other parameters typically depend on the explosion strength. After random sampling, we use these parameters to calculate the desired masses of $^{56}$Ni ($M_{\mathrm{Ni}}$), iron-group elements ($M_{\mathrm{IGE}}$; Ti -- Ni, excluding $^{56}$Ni), intermediate-mass elements ($M_{\mathrm{IME}}$; Ne -- Ca), and unburned material ($M_{\mathrm{C/O}}$) in the ejecta for a given \textsc{tardis} model and scale the abundances accordingly. The time since explosion is randomly sampled from 5 -- 30\,d. The rise time and luminosity at maximum light are used to stretch and scale the predicted light curves of each explosion scenario and determine the luminosity at the requested time. The velocity at maximum light $v_{\textrm{max}}$ is sampled based on existing \textsc{tardis} simulations or observed samples of SNe~Ia and the velocity gradient $\Delta v$ is uniformly sampled from 50 -- 250~~km~s$^{-1}$~d$^{-1}$. These parameters are then used to calculate the inner boundary velocity $v_i$ at the requested time following a linear extrapolation. 

\par

As discussed by \cite{magee--26a}, many parameters are highly correlated or degenerate, but \textsc{riddler} is able to accurately recover the derived parameters (e.g. $M_{\mathrm{Ni}}$, $M_{\mathrm{IGE}}$, $M_{\mathrm{IME}}$, $M_{\mathrm{C/O}}$, $v_i$, etc.) that more directly impact the \textsc{tardis} simulations. We stress however that \textsc{tardis} is insensitive to material below the photosphere and therefore mass estimates from \textsc{riddler} should not be taken as the true masses for a given SN. Instead these are simply the properties of the model that best reproduces the given set of observations. 

\par

For each explosion scenario, 300\,000 synthetic spectra were generated and used to train neural networks that act as \textsc{tardis} emulators. The outputs of the neural networks are the predicted flux and flux uncertainty in 2\,000 log-spaced wavelength bins ranging from 2\,000 -- 10\,000~\AA. Each neural network is trained independently on a single explosion scenario and reaches a typical reduced $\chi^2$ of $\sim$0.8 and prediction error of $\sim$1 -- 3\%.

\par

The neural networks are used in conjunction with \textsc{ultranest} \citep{ultranest} to perform nested sampling and determine the set of parameters $\theta$ that best matches the observations $o$, given the specified priors $\pi(\theta)$ and likelihood function $\mathcal{L}$. During nested sampling, the priors are the same as those used when constructing our \textsc{tardis} models. For a spectral sequence containing $N$ spectra, $\theta$ consists of a set of up to $18 + 2N$ parameters. This includes the 16 input parameters used by the neural networks to generate each spectrum, the distance $d$, host $E(B-V)$ and $R_V$, and a nuisance parameter $f$ designed to account for systematic uncertainties. With the exception of the velocity at maximum light and velocity gradient, all parameters are fixed for all spectra in the sequence. We do however add a constraint that the derived inner boundary velocity decreases over time. 

\par

For the ZTF SN~Ia DR2 sample, we fit for $E(B-V)$ using an exponential prior with $\lambda$ = 0.11 \citep{stanishev--18} and fix $R_V$ = 3.1. We calculate the mean distance modulus based on the redshift, assuming a flat Universe with $H_0 = 67.4$~km~s$^{-1}$~Mpc$^{-1}$ and $\Omega_M$ = 0.315 \citep{planck--20}. The distance modulus uncertainty includes uncertainty from the redshift and a systematic uncertainty of 0.15~mag. 

\par

To compare predicted spectra to observations, we define a Gaussian log-likelihood function that includes the flux uncertainty of both the observations and neural network predictions, and the nuisance parameter $f$. We assume a 5~per~cent flux error for all spectra. \textsc{riddler} also includes the ability to weight each wavelength bin at each epoch independently. Given the photospheric approximation made by \textsc{tardis}, synthetic spectra typically over-predict the flux at longer wavelengths for later epochs. \cite{magee--24} show the impact this approximation has on fitting spectra. Wavelengths $\textgreater$6\,500~\AA\, are therefore excluded from fits at all epochs, but shown here for reference. 

\par

To determine the relative likelihood between two explosion scenarios overall, we use the Bayes factor (BF) given by 
\begin{equation}
\label{eqn:bf}
    \log \mathrm{BF} = \log \mathcal{Z}_1 - \log \mathcal{Z}_2,
\end{equation}
where $Z$ is the evidence calculated by \textsc{ultranest} and given by the marginalised likelihood
\begin{equation}
\label{eqn:evidence}
    \mathcal{Z} = \int \mathcal{L}(\theta|o) \pi(\theta|o) d\theta. 
\end{equation}
A Bayes factor $\textgreater$1 indicates that model 1 is preferred relative to model 2 and BF $\textgreater$5 indicates a strong preference \citep{kaas--95}.

\par

As in \cite{magee--26a}, we stress here that \textsc{riddler} cannot find the `correct' explosion model for a given SN~Ia or determine the overall quality of the fit. Instead, \textsc{riddler} ranks the explosion models from within those covered by the training data in terms of their level of agreement given the priors and likelihood function. This does not preclude the possibility that none of the models are able to match the observations.

%

\section{Results}
\label{sect:results}

\begin{figure*}
    \centering
    \includegraphics[width=\textwidth]{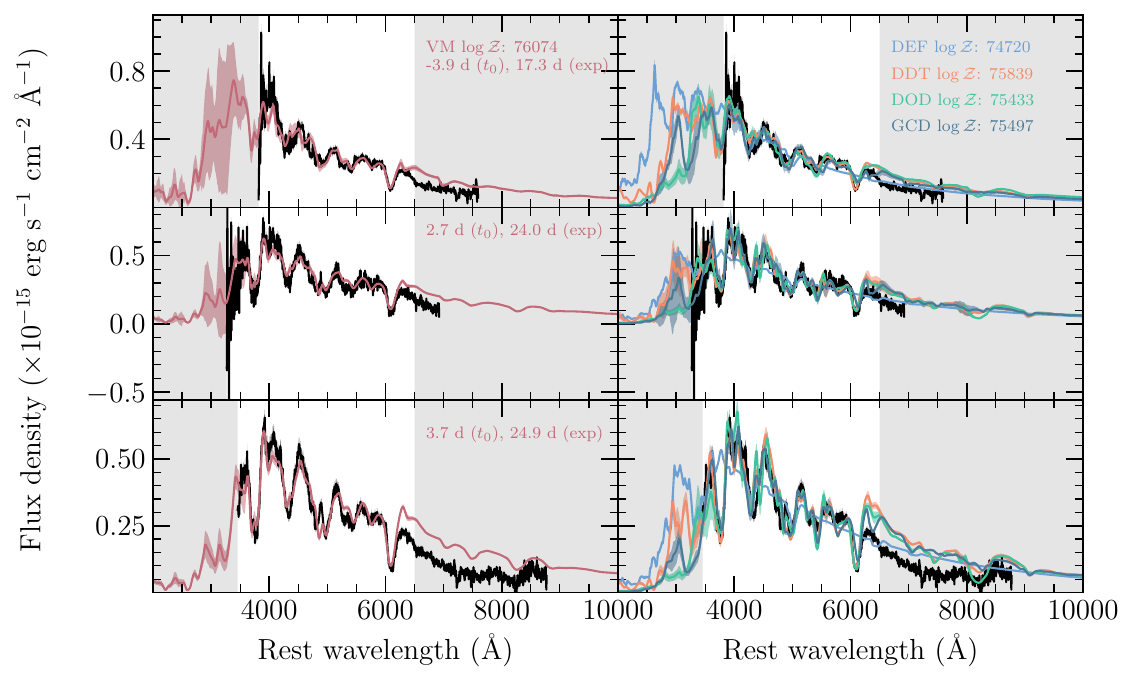}
    \caption{Comparison between spectra of ZTF18aahheaj (black) and emulated spectra produced by \textsc{riddler} for different model types. Emulated spectra are based on the best-fitting model parameters found by \textsc{riddler} for each explosion mechanism. The overall favoured explosion scenario is shown on the left, while all other explosion scenarios are shown on the right. The evidence for each model, determined by \textsc{ultranest}, is shown in the top row. For the best-fitting model, the phase of each spectrum relative to $t_0$ (maximum light) and explosion are given. Shaded grey regions are not included in the fit. Spectra of ZTF18aahheaj have been corrected for Milky Way extinction. The best-fitting host extinction has been applied to emulated spectra. No additional flux offsets or corrections have been applied.}
    \label{fig:ZTF18aahheaj_fit}
\end{figure*}

\begin{figure*}
    \centering
    \includegraphics[width=\textwidth]{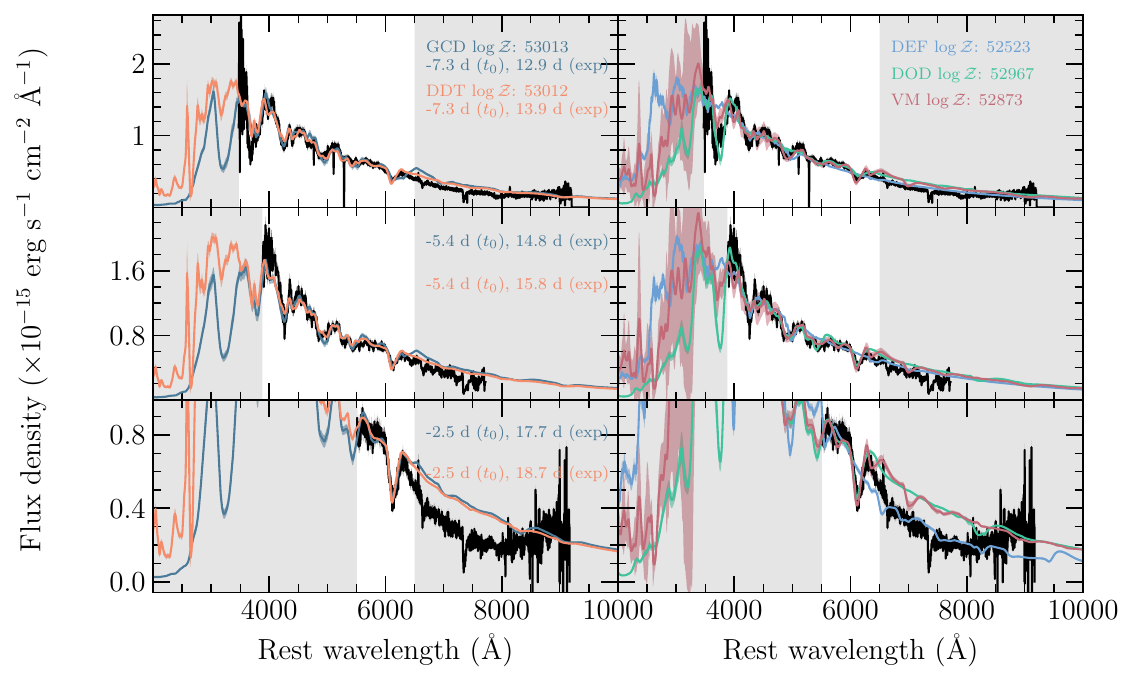}
    \caption{Fits to ZTF18acbxsge as in Fig.~\ref{fig:ZTF18aahheaj_fit}. Both the DDT and GCD models predict similar levels of agreement.}
    \label{fig:ZTF18acbxsge_fit}
\end{figure*}

Figure~\ref{fig:ZTF18aahheaj_fit} shows our fits to ZTF18aahheaj (SN~2018avp), a representative example from our primary sample. Overall, we find that the VM scenario produces the best agreement with ZTF18aahheaj throughout its spectroscopic evolution. The model matches the velocities, widths, and relative strengths of the \ion{Si}{ii}~$\lambda$5\,972 \& $\lambda$6\,355 features at all epochs. The \ion{S}{ii}~W features are also well reproduced, although the bluer feature is somewhat weaker than observed. Around $\sim$5\,000~\AA, ZTF18aahheaj shows a complex blend of \ion{Si}{ii} and \ion{Fe}{ii}. At $-3.9$\,d, the model predicts a single, blended feature around these wavelengths, but by $+2.7$\,d this has developed into two distinct absorption troughs that are in closer agreement with the data. The velocity of the bluer feature however is slightly lower than observed, which is more noticeable for the later $+3.7$\,d spectrum. The VM model also shows good agreement with the \ion{Ca}{ii}~H\&K features at $+2.7$\,d and $+3.7$\,d. Throughout its evolution, our VM model predicts a weak absorption feature due to \ion{C}{ii}~$\lambda$6\,580 that is not clearly observed in the data. In addition, the model flux below $\sim$3\,500~\AA\, is poorly constrained, but we note that these wavelengths are not included in the fit. The DDT model shows a comparable level of agreement at early times, but struggles to reproduce both features around $\sim$5\,000~\AA\, and the \ion{Mg}{ii} and \ion{Si}{iii} absorption at $\sim$4\,200~\AA\, post-maximum. The DDT model also predicts some \ion{C}{ii}~$\lambda$6\,580 that is generally weaker than in the VM model, although still stronger than observed in ZTF18aahheaj. Our DOD model similarly struggles to reproduce the absorption around $\sim$4\,200~\AA\, and predicts strong \ion{Si}{iii}. Pre-maximum, the DOD model shows a lower velocity \ion{Si}{ii}~$\lambda$6\,355 feature than observed. At later epochs the velocity is in closer agreement with ZTF18aahheaj. In contrast to our VM and DDT model, the DOD model matches the blue \ion{S}{ii}~W feature and over-predicts the red absorption. The GCD model generally shows slightly lower \ion{Si}{ii}~$\lambda$6\,355 velocities and stronger \ion{C}{ii}~$\lambda$6\,580 at all epochs. Unsurprisingly, our DEF model is unable to reproduce almost all spectroscopic features and instead predicts a mostly featureless continuum. This likely results from the need to increase the photospheric radius in order to match the luminosity of ZTF18aahheaj.

\begin{figure*}
    \centering
    \includegraphics[width=\textwidth]{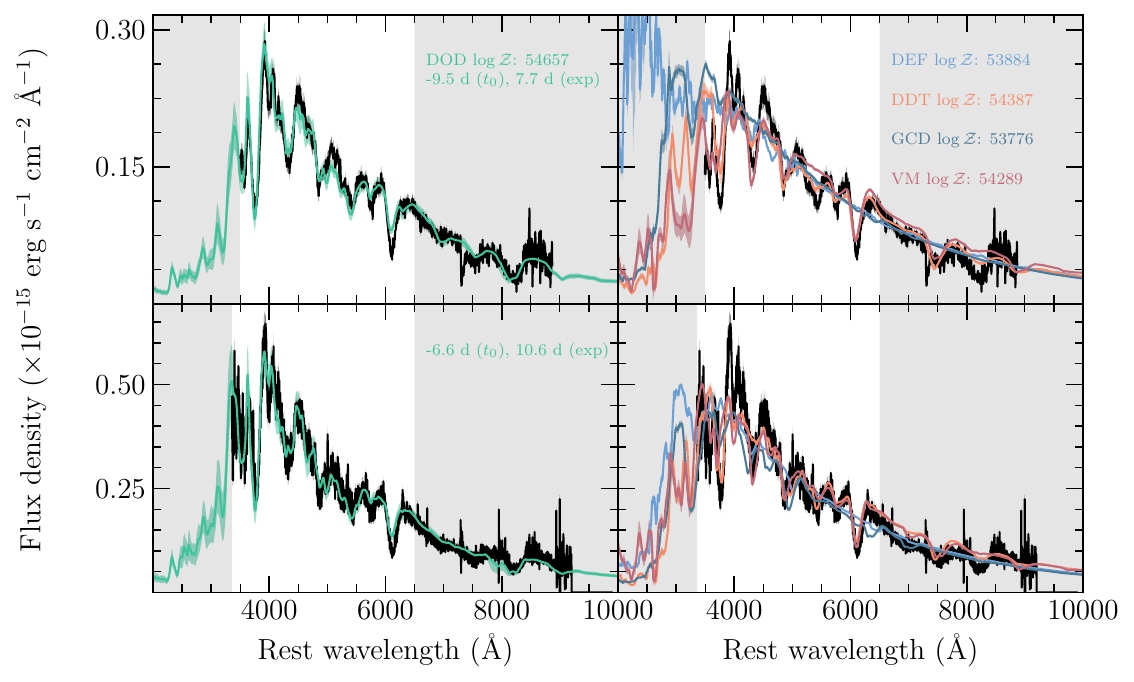}
    \caption{Fits to ZTF19aabvfwn as in Fig.~\ref{fig:ZTF18aahheaj_fit}.}
    \label{fig:ZTF19aabvfwn_fit}
\end{figure*}

\par

The explosion epoch of ZTF18aahheaj can vary considerably depending on the model being fit. For our best-matching model, the VM scenario, we find the earliest explosion epoch of MJD = 59\,203.76$\pm$0.64, while the DEF scenario produces the latest explosion epoch of MJD = 59\,209.21$\pm$0.39. The distance modulii predicted across all models are consistent within their uncertainties (VM $\mu = 36.99\pm0.05$) and all models predict relatively minor host extinction (host $A_V \lesssim 0.05$). For our VM scenario, we find that the ejecta model is comparable to that predicted by the merger of a $\sim$1.1~$M_{\odot}$ primary white dwarf and a $\sim$0.90~$M_{\odot}$ secondary, which contains $0.56\pm0.02~M_{\odot}$ of $^{56}$Ni, $0.31\pm0.01~M_{\odot}$ of IMEs, and the remaining material is unburned carbon and oxygen\footnote{As noted by \cite{magee--26a}, we stress that the abundances measured in this work should not be taken as literal measurements of abundances in individual SNe. \textsc{tardis} is insensitive to material below the photosphere and therefore cannot constrain the total masses of different elements. Values reported here are given as reference for the properties of the best-fitting models.}. All models predict $^{56}$Ni masses ranging from $\sim$0.4 -- 0.6~$M_{\odot}$. We note that the total ejecta mass of our VM model ($\sim2~M_{\odot}$) is at the upper limit of our training data. This could explain the relatively large uncertainties of the model in the UV, particularly for the first spectrum, as the neural networks may have struggled to learn due to the large variations expected at these wavelengths and the limited training data available.

\par

For some SNe, we find similar levels of agreement between multiple models and therefore are unable to determine a single best-matching explosion scenario. An example of this is shown in Fig.~\ref{fig:ZTF18acbxsge_fit} for the case of ZTF18acbxsge (SN~2018huw). Here we find that the DDT and GCD scenarios both produce comparable spectra across the three observed epochs despite the differences in their ejecta structures. At $-7.3$\,d, both models generally reproduce the overall shape of ZTF18acbxsge and most of the spectral features. The most prominent difference is the \ion{Si}{ii}~$\lambda$6\,355 feature, for which our GCD model shows a much broader component towards the red wing. This may be due to additional absorption from high velocity \ion{C}{ii}~$\lambda$6\,580 produced by the deflagration ash. Indeed, some \ion{C}{ii} absorption is also produced in the GCD model at later epochs. The DDT model does not predict strong \ion{C}{ii} features, in better agreement with ZTF18acbxsge. At $-5.4$\,d and $-2.5$\,d, the DDT and GCD models are again very similar with only minor differences in the strengths of some features. Despite these similarities in the optical, Fig.~\ref{fig:ZTF18acbxsge_fit} shows that both models predict very different UV spectra and therefore UV spectra can be a powerful tool for discriminating between explosion scenarios. In particular, the DDT model shows a bluer colour and relatively weak \ion{Ca}{ii}~H\&K absorption throughout its evolution, while the \ion{Ca}{ii} absorption in the GCD model is significantly stronger. Unfortunately, none of the spectra of ZTF18acbxsge extend far enough to blue wavelengths to discriminate between these two cases. Both the DDT and GCD models predict similar explosion epochs (MJD = $58\,418.63\pm0.56$ and $58\,419.64\pm0.47$, respectively), distance modulii ($\mu = 36.12\pm0.06$ and $35.81\pm0.09$, respectively), and negligible host extinction. Despite similar distance modulii, the DDT model contains a larger $^{56}$Ni mass of $1.03\pm0.05~M_{\odot}$ compared to $0.76\pm0.05~M_{\odot}$ for the GCD model.

\par

In Fig.~\ref{fig:ZTF19aabvfwn_fit}, we show that ZTF19aabvfwn (SN~2019jf) is best fit by our DOD model. The model predicts somewhat weaker \ion{Si}{ii}~$\lambda$6\,355 than observed at all epochs, but generally is able to reproduce most of the spectral features and continuum shape. In addition, Fig.~\ref{fig:ZTF19aabvfwn_fit} shows that even though wavelengths $\textgreater$6\,500~\AA\, are not included in our fits, the DOD scenario is able to broadly match the continuum and \ion{Ca}{ii}~NIR triplet at these epochs. The DOD model ejecta is comparable to the detonation of a 0.96~$M_{\odot}$ white dwarf core with a 0.07~$M_{\odot}$ helium shell. The majority of the material in the shell is unburned, with only a small amount of IGEs ($\sim0.001~M_{\odot}$) and IMEs ($\sim0.008~M_{\odot}$).

\par

Fits to the remainder of our primary sample are shown in Appendix~\ref{sect:primary_figures}, while details of the fits are given in Appendix~\ref{sect:primary_tables}. We note that predicted spectra for some SNe~Ia show relatively large uncertainties. For example, our fits to ZTF18abosdwf (SN~2018fcw; Fig.~\ref{fig:ZTF18abosdwf_fit}) show typical uncertainties ranging from $\sim$1 -- 20~per~cent depending on the model. In this case, the relatively large uncertainties may arise from issues with the data. Our fits include two spectra of ZTF18abosdwf taken with different telescopes on the same night. Differences in noise properties between the two spectra could mean that our fits struggle to find a model that works equally well for two different spectra at the same epoch. This could result in the fits pushing to regions of the parameter space that are poorly sampled, in an effort to increase the uncertainties such that they cover both observed spectra. Alternatively, the best-fitting parameter values may be intrinsic to ZTF18abosdwf and therefore it is not-well covered by our training data. Similarly, the best-fitting parameters for ZTF18aaumeys (SN~2018bpd; Fig.~\ref{fig:ZTF18aaumeys_fit}) result in spectra with slightly larger uncertainties of $\sim$2 -- 40~per~cent. Again this could indicate that ZTF18aaumeys is not well-covered by our training data or issues with the data, specifically the flux calibration. The SALT2 model struggles to reproduce the $i$-band light curve, which will affect the continuum shape of our final spectra. Hence, there may be a mismatch between the continuum shapes and spectral features that our models cannot reproduce.

\par

Table~\ref{tab:best_fitting_types} gives the number of SNe~Ia in each sample that are best reproduced by our different explosion models. For our primary sample, we find 11 ($\sim39$~per~cent) SNe~Ia are best matched by Chandrasekhar mass explosion scenarios. Of these, 6 favour DDT models and 3 favour GCD models. For 2 SNe, we find similar levels of agreement between both explosion scenarios and therefore are unable to distinguish between them. The majority of the primary sample ($\sim61$~per~cent) are best matched by sub-Chandrasekhar mass models, with 10 SNe favouring DOD models and 7 favouring VM models. Interestingly, we find a similar breakdown between Chandrasekhar mass and sub-Chandrasekhar mass models for our secondary sample. In both samples, approximately one third of SNe~Ia favour Chandrasekhar mass models, while two thirds favour sub-Chandrasekhar mass models. Although the limitations and biases of our sample prevent us from determining an intrinsic rate of each explosion or progenitor type, our results highlight the fact that multiple production channels for SNe~Ia may indeed be found in cosmological samples.

\par

\begin{table}
\begin{center}
\caption{Best-fitting model types to the ZTF SN~Ia DR2 sample.}
\label{tab:best_fitting_types}
\begin{tabular}{lll}
\hline
\textbf{Model type} & \textbf{Primary} & \textbf{Secondary}  \\
 & \textbf{sample} & \textbf{sample}  \\
\hline
\hline
\textbf{Chandrasekhar mass} & \textbf{11} & \textbf{79}\\
~~~~DEF & 0 & 2 \\
~~~~DDT & 6 & 48 \\
~~~~GCD & 3 & 26 \\
~~~~DDT/GCD & 2 & 3 \\
\textbf{sub-Chandrasekhar mass} & \textbf{17} & \textbf{130}\\
~~~~DOD & 10 & 90 \\
~~~~VM  & 7  & 36 \\
~~~~DOD/VM & 0 & 4 \\
\textbf{Unclear} & \textbf{0} & \textbf{7} \\
\hline
\textbf{Total} & \textbf{28} & \textbf{216} \\
\hline
\hline
\end{tabular}
\end{center}
\end{table}

Surprisingly, we find that two SNe~Ia in our secondary sample, ZTF19aafmfxg (SN~2019ahr; Fig.~\ref{fig:ZTF19aafmfxg_fit}) and ZTF19acxgxcu (SN~2019weu; Fig.~\ref{fig:ZTF19acxgxcu_fit}), are best matched by DEF models. Such models are typically associated with peculiar SNe~Iax \citep{fink-2014}, due to the low $^{56}$Ni masses and velocities they predict, and therefore would not be expected to appear in samples of cosmological SNe~Ia. ZTF19aafmfxg was observed spectroscopically approximately one week prior to maximum light and is classified in the ZTF SN~Ia DR2 as a 91T-like SN. At these early epochs, spectra of SNe~Iax show strong similarities to 91T-likes, including blue colours and prominent \ion{Fe}{iii} absorption \citep{02cx--orig}. These features are well reproduced by our DEF model, although the model does not match the weak \ion{Si}{ii}~$\lambda$6\,355 absorption. In contrast all other models tend to over-predict the strength of the \ion{Si}{ii}~$\lambda$6\,355 absorption, in disagreement with the observations. Based on this fit, we find an explosion epoch that falls against the upper limit of our prior (MJD $\sim$ 58\,511.63). All other models predict earlier explosion epochs by at least four days and comfortably within our prior range. Similarly, with our DEF model we find a distance modulus $\sim4\sigma$ below the value estimated from the redshift and assuming a 0.15~mag systematic uncertainty. All other models predict distance modulii within $\sim1\sigma$ of the redshift estimated value. ZTF19aafmfxg is classified as a peculiar, 02es-like SN~Ia \citep{ganeshalingam--12}. Similar to SNe~Iax, 02es-like SNe are sub-luminous and some members of the class also exhibit low ejecta velocities. ZTF19aafmfxg was observed at $+4.0$\,d and $+12.9$\,d, but due to the limits of our training data only the first spectrum is included in our fits. Although not included, we note that the $+12.9$\,d spectrum shows strong similarities to the prototypical SN~Iax, SN~2005hk \citep{phillips--07}, approximately two weeks after maximum light. While our DEF model generally reproduces features $\lesssim$5\,000~\AA, it struggles to match longer wavelengths. \cite{magee--26a} report a similar issue when fitting a spectrum of SN~2005hk around this epoch, indicating it may be a systematic issue with our DEF training data at these phases. Our other models show similar difficulties in reproducing the spectra and predict strong \ion{Si}{ii}~$\lambda$6\,355 at higher velocities than observed. Again we find that the explosion epoch predicted by our DEF model falls against our assumed upper limit on the prior (MJD $\sim$ 58\,819.52). In this case however the larger distance modulus uncertainty (the redshift of ZTF19aafmfxg is estimated from \textsc{snid} template fits and therefore is somewhat larger) means that our best-fitting distance modulus from the DEF model is within $\sim2\sigma$ of the redshift-estimated value. Therefore while tighter constraints on the priors and/or additional spectra may find alternative models provide better agreement with these SNe, our fits present plausible evidence of deflagration explosions reproducing some peculiar SNe~Ia that may make their way into cosmology samples.

%

\section{Discussion}
\label{sect:discussion}

\subsection{Biases}
\label{sect:biases}

\begin{figure}
\centering
\includegraphics[width=\columnwidth]{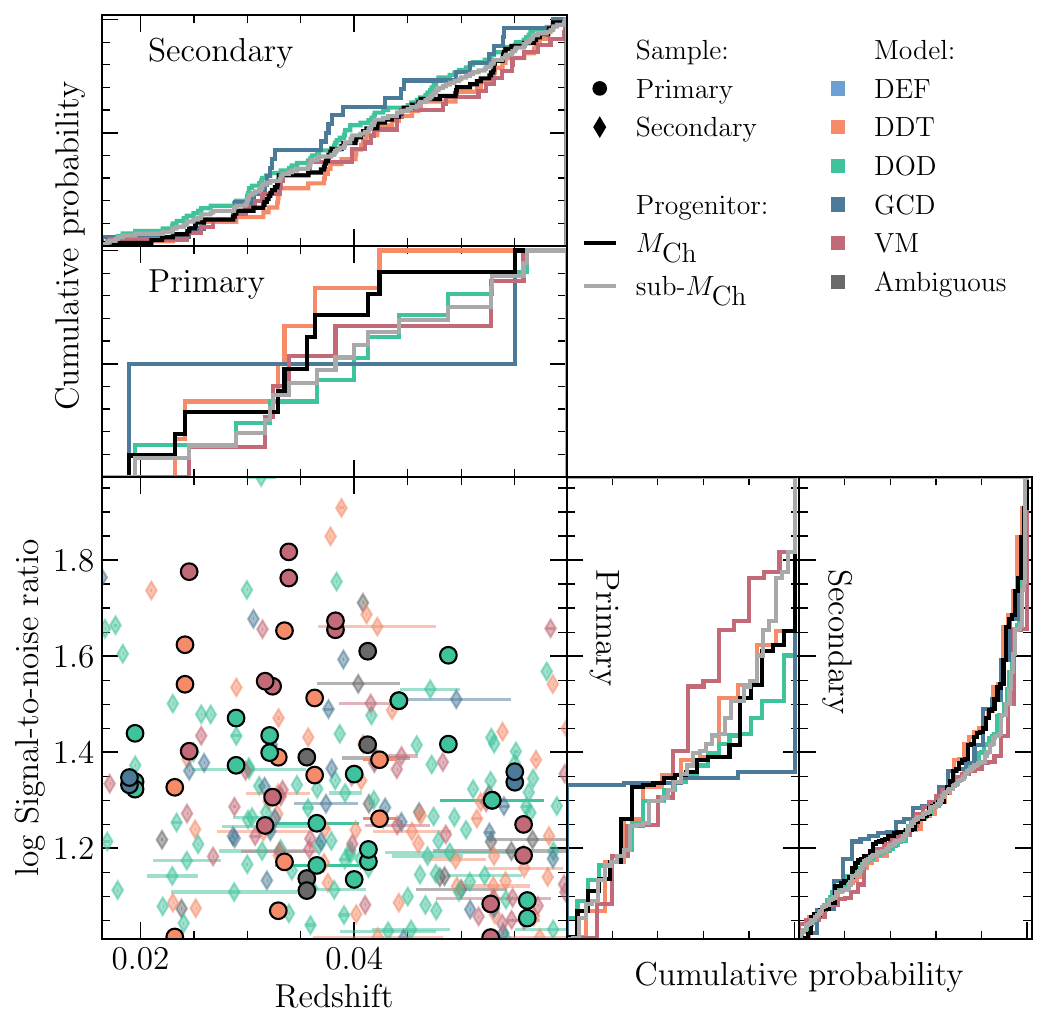}
\caption{Distributions of log signal-to-noise ratios and redshifts of SNe~Ia in our primary and secondary samples. Circles show our primary sample, while diamonds show our secondary sample. Each point shows a spectrum and is coloured based on the overall best-fitting explosion scenario for that SN. Note that SNe~Ia in our primary sample have multiple observations and therefore each SN is shown multiple times. Top and right-hand panels show cumulative distribution functions for properties depending on the explosion model or progenitor type for our primary and secondary samples separately.
}
\label{fig:bias_plots_snr_z}
\centering
\end{figure}

As discussed in Sect.~\ref{sect:sample}, our samples are inherently biased. Here we explore whether properties of the SNe or spectra themselves could also introduce a bias that affects which explosion scenario is favoured or the properties of the best-fitting model.

\par

\begin{figure}
\centering
\includegraphics[width=\columnwidth]{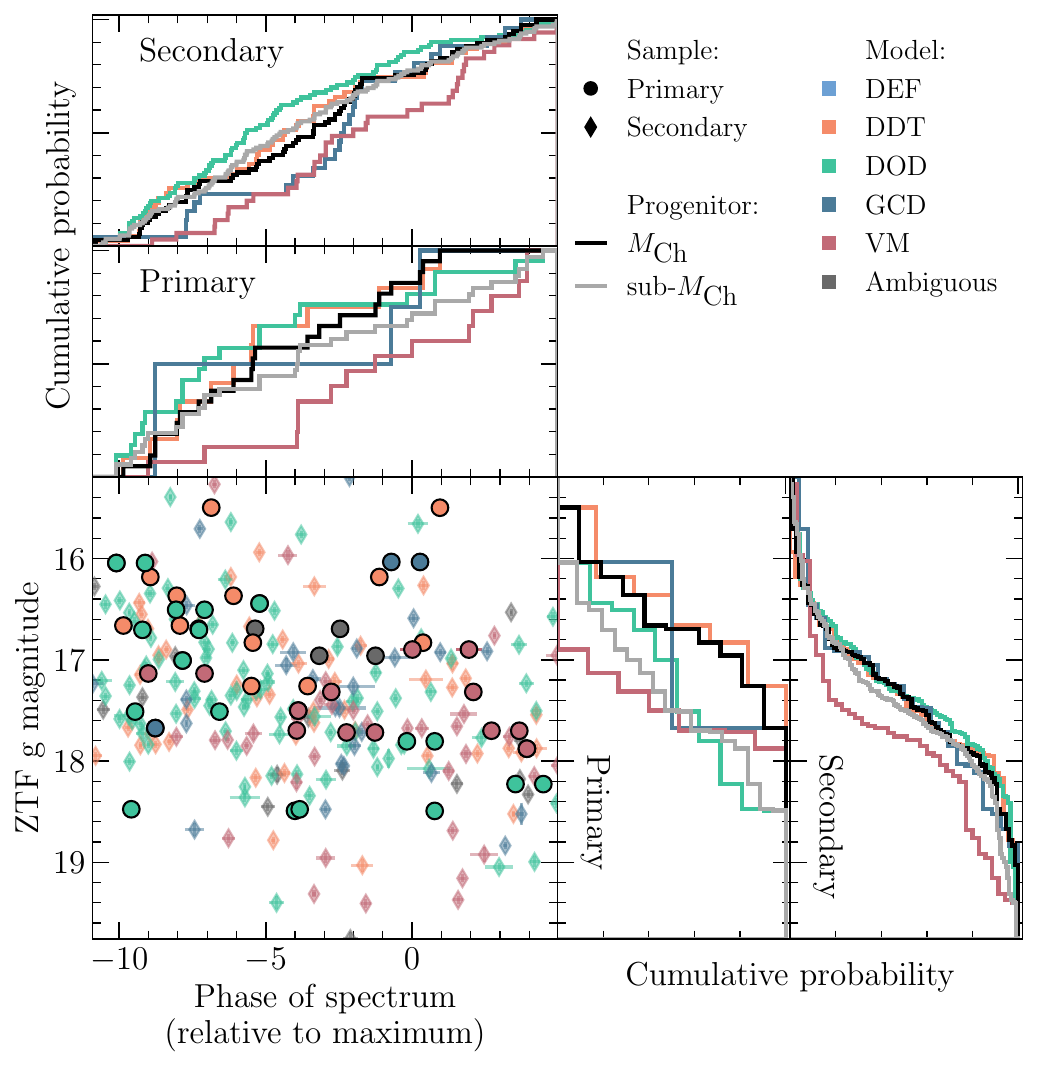}
\caption{As in Fig.~\ref{fig:bias_plots_snr_z} for the SALT2 model ZTF $g$-band magnitude at the phase of the spectrum and the phase of each spectrum.
}
\label{fig:bias_plots_mag_phase}
\centering
\end{figure}

Figure~\ref{fig:bias_plots_snr_z} shows the distributions of signal-to-noise ratios and redshifts for the SNe~Ia in our primary and secondary samples. As indicated by Fig.~\ref{fig:bias_plots_snr_z}, the redshift distribution of DDT models in our primary sample appears to show some bias towards lower redshifts, although this is based on only six SNe~Ia and not statistically significant. While all other explosion scenarios are found across the full redshift range of our primary sample, $z \sim$ 0.02 -- 0.06, DDT models are not favoured for any SNe~Ia above $z \gtrsim 0.04$. This could indicate that DDT models are perhaps marginally favoured for brighter, and hence more nearby, SNe and there should exist a correlation with signal-to-ratio. Indeed, Fig.~\ref{fig:bias_plots_mag_phase} shows that DDT models are not favoured for any spectra in our primary sample with ZTF $g$-band apparent magnitudes fainter than $\sim$17.5. As demonstrated by Fig.~\ref{fig:bias_plots_snr_z} however, there is no clear preference for DDT models in spectra with higher signal-to-noise ratios -- all model types are found across the full range of signal-to-noise ratios examined here. Similarly, our secondary sample also shows that some DDT models are in fact favoured at higher redshifts and fainter magnitudes. Conversely, the VM models in our primary sample are found across all redshifts, but do appear to display a slight preference for higher signal-to-noise ratios compared to other models (again this is not statistically significant). This is despite the fact that our VM models are also generally favoured for fainter SNe, with no VM models favoured for any SNe with spectra brighter than $\sim$16.8 (Fig.~\ref{fig:bias_plots_mag_phase}). We note that VM models in our secondary sample show no clear preference for higher signal-to-noise ratios.

\par

Rather than a bias towards different magnitudes or signal-to-noise ratios, the trends observed among our DDT and VM models may instead be secondary effects arising from biases towards different phases. Figure~\ref{fig:bias_plots_mag_phase} shows that DDT models are somewhat favoured for earlier spectra, while VM models are somewhat favoured for later spectra within both our primary and secondary samples. The DDT models upon which the \textsc{riddler} training data is based could provide a better match to the outer layers (and hence early spectra) of (cosmological) SNe~Ia. Likewise, the VM models may provide a better match to the inner layers (and later spectra). Figure~\ref{fig:bias_plots_mag_phase} also shows a negative correlation between magnitude and phase (spectra taken at later phases are fainter) that is counter to what one may naively expect from SN evolution (SNe get brighter towards maximum light). This correlation further highlights a spectroscopic selection effect for our sample and helps to explain the trends observed among DDT and VM models. SNe~Ia that are more nearby are typically brighter at earlier phases and therefore can be observed more easily, hence DDT models favour lower redshifts. At the same time, fainter SNe can only be observed closer to maximum light and may be preferentially selected for larger telescopes, hence VM models favour higher signal-to-noise ratio spectra despite being fainter. Determining which scenario is favoured overall for a given SN would require multiple spectra covering these phases and this could result in changes to the best-fitting model compared to those found here. Our overall results may therefore be biased due to a complicated and uneven phase distribution and spectroscopic selection effects for our samples. Removing such a bias would require each SN to have spectra at a consistent set of phases from pre- to post-maximum light.

\par

\begin{figure}
\centering
\includegraphics[width=\columnwidth]{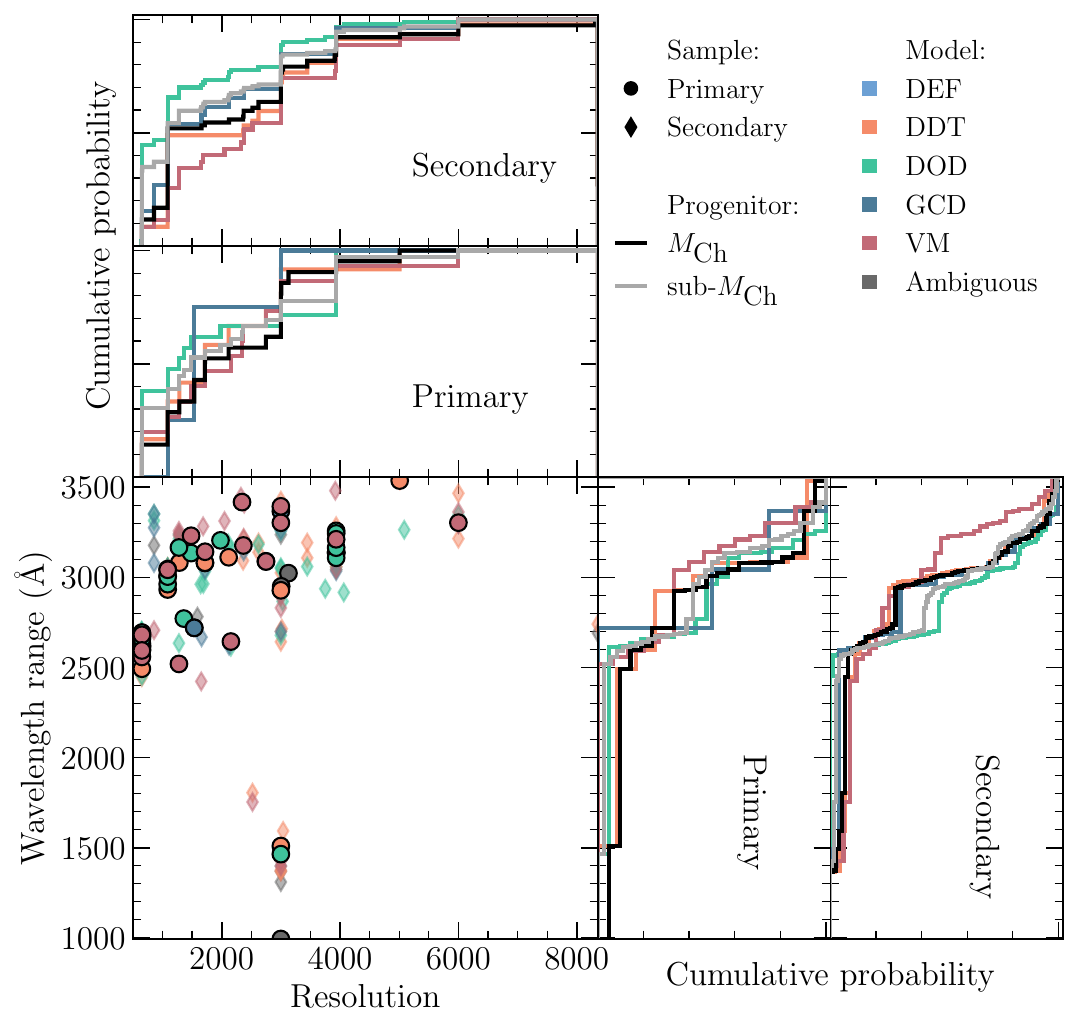}
\caption{As in Fig.~\ref{fig:bias_plots_snr_z} for the wavelength range and resolution of each spectrum. 
}
\label{fig:bias_plots_wave}
\centering
\end{figure}

\begin{figure*}
\centering
\includegraphics[width=0.7\textwidth]{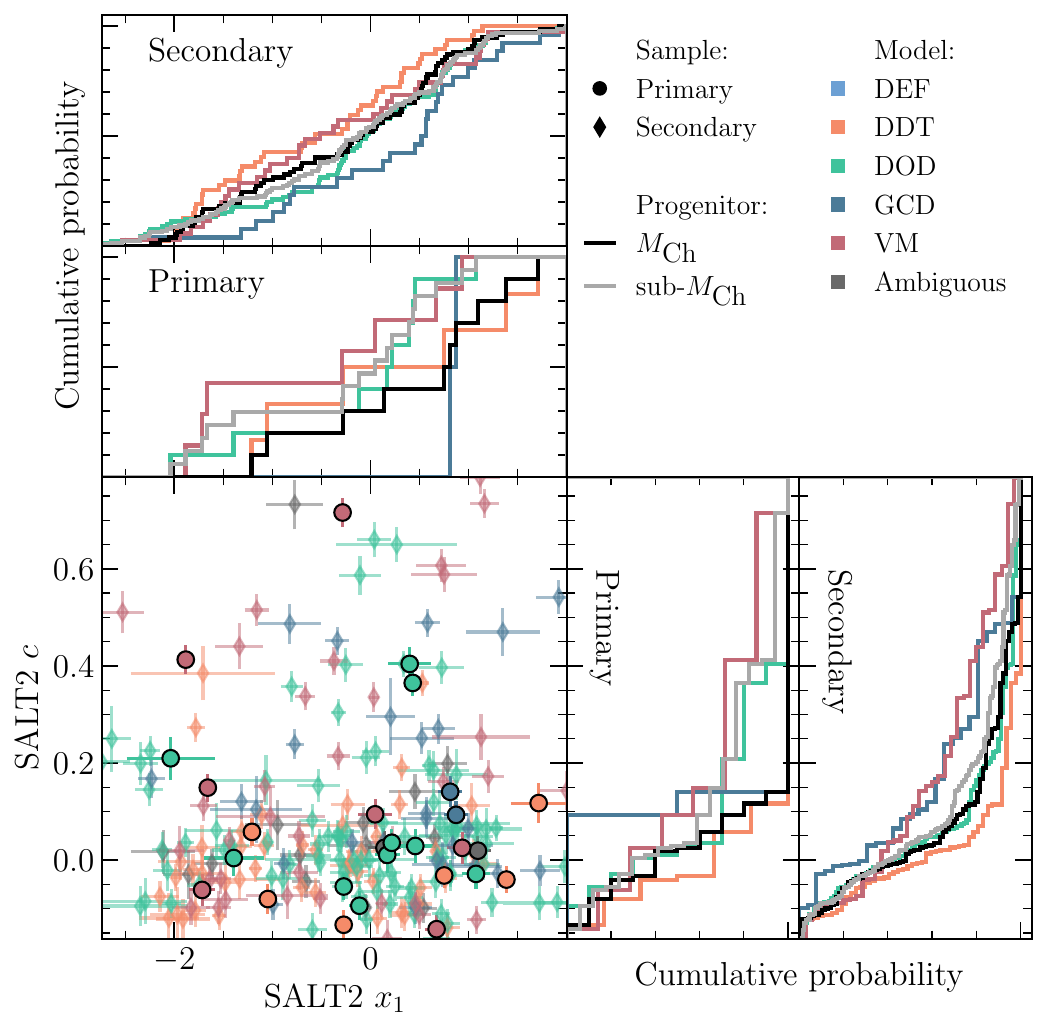}
\caption{Distributions of SALT2 $x_1$ and $c$ parameters for our primary and secondary samples. Circles show our primary sample, while diamonds and faded lines show our secondary sample. Each point is coloured based on the overall best-fitting explosion scenario. 
}
\label{fig:salt}
\centering
\end{figure*}

Finally, in Fig.~\ref{fig:bias_plots_wave} we show the distributions of spectral properties for our samples. During our fits, we do not include wavelengths beyond $\lambda$ \textgreater 6\,500~\AA, therefore the wavelength range shown in Fig.~\ref{fig:bias_plots_wave} effectively measures how far into bluer wavelengths the fit extends. We find no difference in the wavelength range distributions of our various models. For our primary sample, we find no significant differences between the resolution distributions of our models. Interestingly however, for our secondary sample, we find that the DOD scenario appears to be marginally favoured for lower resolutions compared to other scenarios. While this may be related to intrinsic differences in the velocities and shapes of spectral features between the different model scenarios, as discussed in Sect.~\ref{sect:sample}, \cite{magee--26a} find that resolutions above $\gtrsim$300 do not significantly impact the results of fits.

\subsection{SALT2 model parameters}
\label{sect:salt}

In Fig.~\ref{fig:salt} we show the distributions of SALT2 $x_1$ and $c$ parameters for each model type, and the correlations between them. Based on our results, we find hints that $x_1$ may be somewhat correlated with the mass of the exploding white dwarf. Grouping our models into explosions of Chandrasekhar and sub-Chandrasekhar mass progenitors, within our primary sample the former shows some indication of a systematic trend towards higher $x_1$. DDT models are not favoured for the fastest evolving SNe~Ia ($x_1 \lesssim -1.2$) and GCD models are favoured only for the slowest evolving SNe~Ia ($x_1 \gtrsim 0.8$). Conversely, the DOD and VM models are generally spread across $x_1$, although DOD models also appear to favour higher $x_1$ compared to VM models. We caution however that this is based on only a small number of objects (e.g. 6 DDT and 2 GCD models) and therefore is not statistically significant. Verifying this trend would require a larger sample. Indeed, our secondary sample does not appear to show any significant differences between Chandrasekhar and sub-Chandrasekhar mass progenitors, but we again note that these fits may not be robust (see Sect.~\ref{sect:sample}).

\par

Although our models are insensitive to material below the photosphere and therefore cannot measure a true ejecta (or $^{56}$Ni) mass, the spread and overlap in different explosion scenarios across $x_1$ would imply that we also do not find a strong correlation between light curve shape and ejecta mass. Previous studies invoking analytical models to fit SNe~Ia light curves have found correlations between $x_1$ and both ejecta mass and $^{56}$Ni mass \citep{scalzo--14c, scalzo--19, sarin--26}. In contrast however, explosion simulations predict a more complicated picture. The delayed detonation models presented by \cite{blondin--13} all contain $\sim$1.4~$M_{\odot}$ of ejecta, but show rise times that vary by $\sim$5\,d. Even more extreme, in the helium-ignited violent merger scenario the secondary white dwarf may also undergo a double detonation explosion. In this case, the total ejecta mass of the system is increased by a factor of up to $\sim$2 and yet the light curves and spectra show only minor differences during phases up to and shortly after maximum light \citep{pakmor--22, boos--24, pollin--24}. The limitations of our sample prevent us from fully investigating correlations between light curve shape and other properties of the explosion, but our results generally support the conclusion that the light curve shape is unlikely to be solely driven by the ejecta mass.

\par

\begin{figure}
\centering
\includegraphics[width=\columnwidth]{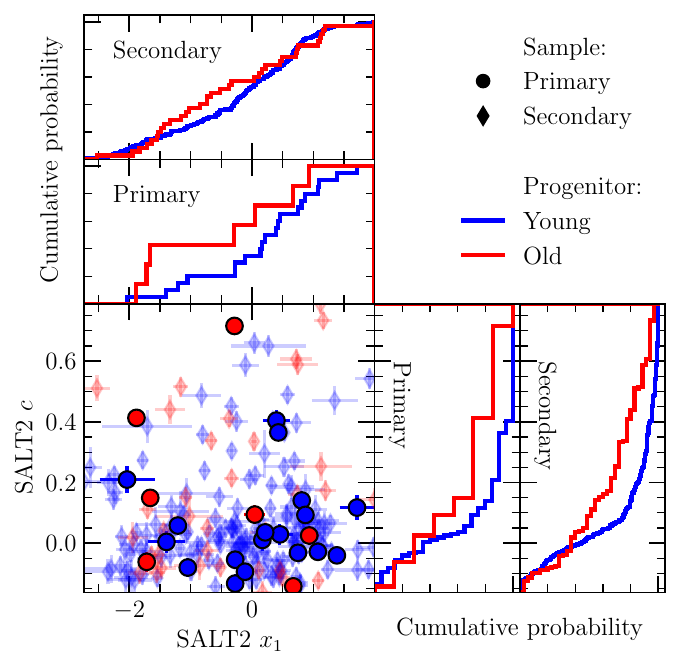}
\caption{As in Fig.~\ref{fig:salt}, but assuming all VM models arise from older populations and all other explosion scenarios arise from young populations. 
}
\label{fig:salt_prog}
\centering
\end{figure}

Rather than grouping by the mass of the exploding progenitor or ejecta, we can attempt to group our models by progenitor age (Fig.~\ref{fig:salt_prog}). The $x_1$ distribution of SNe~Ia has been shown to be non-Gaussian (e.g. \citealt{scolnic--16, jones--19, ginolin--25b}). Typically this is modelled as a population of young progenitors with high $x_1$ values and a population of older progenitors showing both high and low $x_1$ \citep{rigault--20}. The double degenerate progenitor scenario predicts a broad range of delay times between star formation and explosion, while single degenerate scenarios are typically limited to shorter delay times (within a few Gyr; \citealt{ruiter--09, ruiter--11, maoz--10, liu--23--rev}). Therefore, assuming that our VM models represent an old population and all other models combined represent a young population, our results are qualitatively consistent with this interpretation -- older progenitors are found across all $x_1$ and younger progenitors are skewed towards higher $x_1$. This is purely speculative however as it is based on assuming all VM SNe~Ia arise from old progenitors, whereas in practice there will be at least some overlap between the delay times of the different explosion scenarios. In addition, there are significant uncertainties associated with theoretical predictions of delay times from binary population synthesis calculations \citep{wang--12, maoz--14, toonen--14}. Finally, as previously noted, our sample is heavily biased and therefore the distributions of $x_1$ found for different explosion scenarios do not necessarily reflect their intrinsic distributions. Nevertheless, our results highlight that multiple progenitor and/or explosion scenarios could plausibly be identified within cosmological SNe~Ia samples. While certain progenitor and/or explosion scenarios may show systematic biases towards higher or lower values, they are not easily distinguished based purely on $x_1$. 

\par

Figure~\ref{fig:salt} also shows the distribution of SALT2 $c$ values. As with $x_1$, we find some indications of differences in the distributions of $c$ for Chandrasekhar and sub-Chandrasekhar mass progenitors, but these are again not statistically significant. Within our primary sample, DDT and GCD models are favoured only for SNe~Ia with relatively blue colours ($c \lesssim$ 0.15), while our DOD and VM models extend to much redder colours. Red colours are generally a prediction of double detonation explosions invoking large helium shell masses \citep{kromer--10, collins--22}. In this scenario, nuclear burning within the helium shell results in the production of significant amounts of iron-group elements (IGEs) in the outer ejecta. When viewed close to the point at which the shell was ignited, these IGEs lead to strong line-blanketing and redder colours than typically observed for normal SNe~Ia. Viewed away from the ignition point, these effects become less pronounced and lead to colours closer to normal SNe~Ia. Our DOD models could therefore be biased towards those systems viewed off-axis -- the most extreme cases viewed close to the ignition point are unlikely to pass cosmology cuts and therefore would have been removed from our sample. Alternatively, lower helium shell masses or shells polluted with $^{14}$N also produce fewer IGEs in the outer ejecta and predict colours that are more typical of normal SNe~Ia \citep{shen--14b, polin--19, townsley--19}. The fact that we find some blue DOD models could instead suggest that thin and/or polluted helium shells are relatively common among cosmological SNe~Ia. Indeed, \cite{townsley--19} show that only a modest amount of $^{14}$N is required ($\lesssim1$~per~cent) and some $^{14}$N is expected either from accretion or previous phases of helium burning \citep{shen--09}.

\par

\begin{figure}
\centering
\includegraphics[width=\columnwidth]{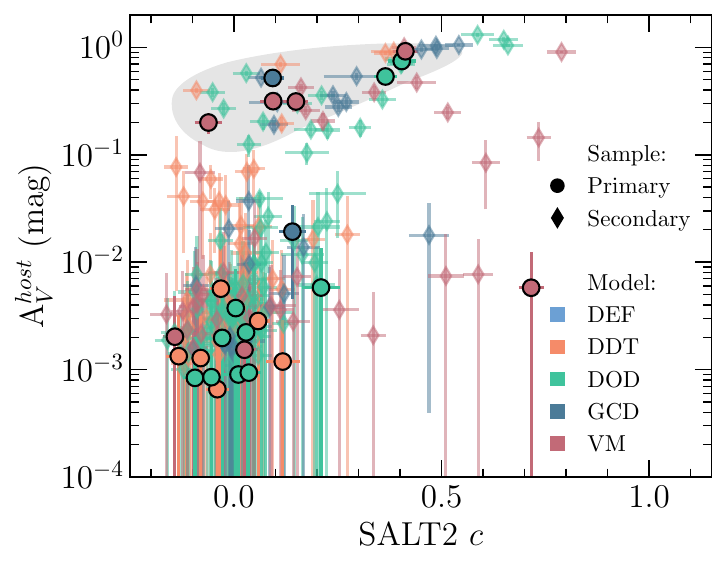}
\caption{Comparison between SALT2 $c$ and host extinction estimated from our best-fitting \textsc{riddler} models. Uncertainties on $A_V^{host}$ are given as 1$\sigma$ errors. The grey shaded region approximately shows intrinsically blue SNe~Ia reddened by dust. We find a number of outliers with high $c$ and low $A_V^{host}$ that may be a result of our prior assumptions.
}
\label{fig:extinction}
\centering
\end{figure}

Some SNe~Ia light curve fitters attempt to separate the effects of intrinsic colour and extrinsic dust extinction \citep{jha--07, burns--11, mandel--22}. The SALT2 $c$ parameter however is an empirical property that combines the two into a single variable. In Fig.~\ref{fig:extinction} we show the inferred host extinction from our \textsc{riddler} fits compared against the SALT2 colour. Unsurprisingly, we find a correlation between the two -- higher $A_V^{host}$ results in higher SALT2 $c$. Based on light curve fits to the ZTF SN~Ia DR2 sample using a phenomenological model, \cite{sarin--26} report a similar, although somewhat stronger, correlation with a Spearman correlation coefficient of $\rho = 0.83$. Here we find $\rho \sim 0.6$. Dust extinction and SALT2 $c$ may be correlated, but the scatter in our results highlight that extinction is not the sole property driving $c$ variation and the intrinsic colour also plays an important role. Similar scatter was also observed by \cite{sarin--26}, particularly for $c \lesssim 0.5$. 

\par

\begin{figure*}
\centering
\includegraphics[width=0.7\textwidth]{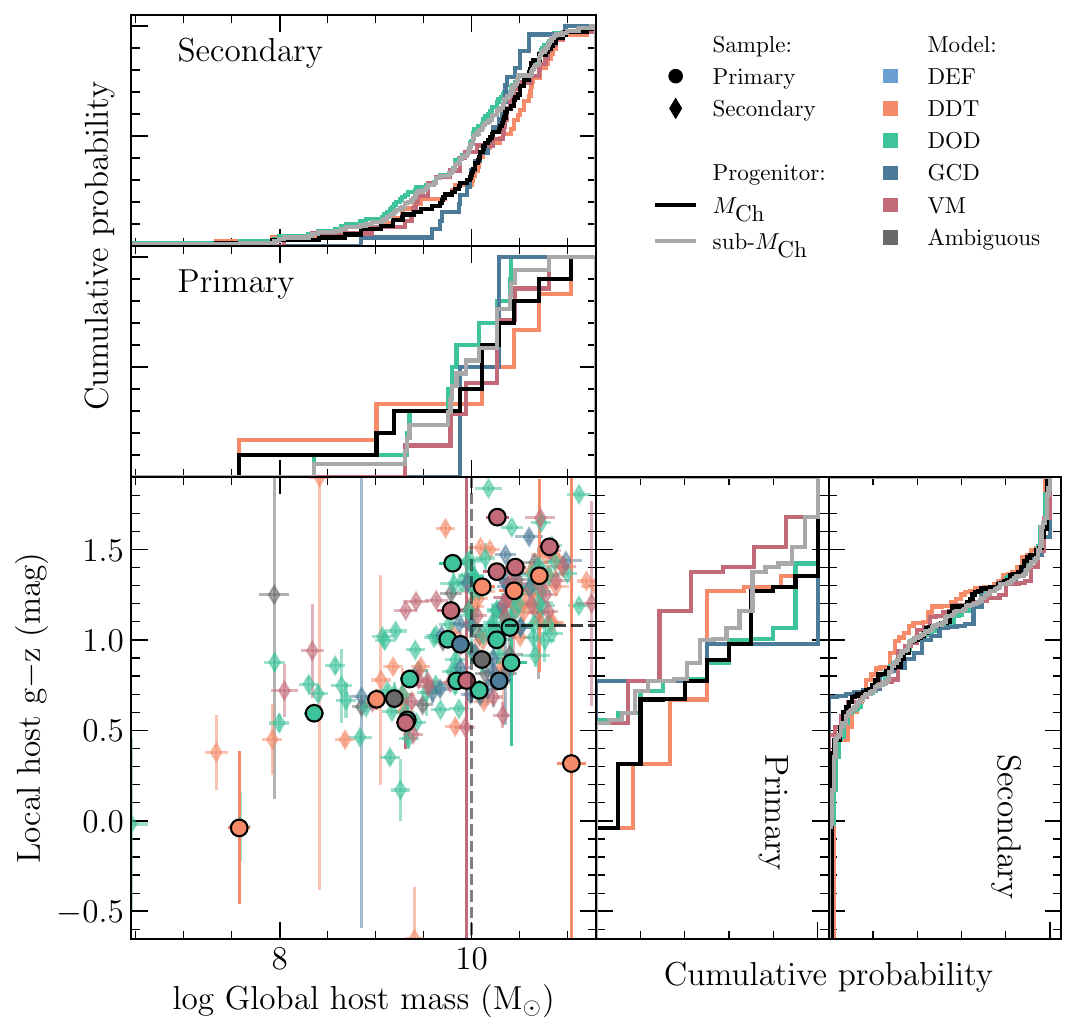}
\caption{As in Fig.~\ref{fig:salt} for global host galaxy mass and local host galaxy restframe $g-z$ colour. Dashed lines in the bottom left panel give indicative regions separating low-mass, high-mass star-forming, and high-mass passive galaxies. 
}
\label{fig:host}
\centering
\end{figure*}

Figure~\ref{fig:extinction} shows that SNe~Ia can be broadly separated into at least two sub-groups based on their $A_V^{host}$ and $c$. The bulk of the population experiences minimal host extinction ($A_V^{host} \lesssim 0.1$) and shows relatively blue colours ($c \lesssim 0.1$). The second sub-group (shaded region) represents intrinsically blue SNe~Ia that have been reddened by dust. These SNe~Ia have SALT2 colours spanning the range of our sample and $A_V^{host} \gtrsim 0.1$. An example of this is ZTF19acazbqm (SN~2019roz; Fig.~\ref{fig:ZTF19acazbqm_fit}) with $c = 0.15\pm0.03$ and for which we find the best agreement with a VM model and $A_V^{host} = 0.31\pm0.05$. Spectra of ZTF19acazbqm show a sharp absorption feature consistent with \ion{Na}{i}~D and indicative of dust extinction (e.g. \citealt{na1d-red-1, na1d-red-2}). Similarly, for ZTF19aamkdtq (SN~2019bsa; Fig.~\ref{fig:ZTF19aamkdtq_fit}) $c = -0.06\pm0.03$ and we find the best agreement with a VM model and $A_V^{host} = 0.20\pm0.04$. The first SNIFS \citep{aldering--02} spectrum at maximum light clearly shows a large gap where the blue and red arms have been joined. The resulting spectrum is somewhat redder than expected from the ZTF photometry and the level of host extinction may be slightly overestimated. The colour of the second spectrum taken approximately two days later however is well matched by our model. Both spectra also show \ion{Na}{i}~D absorption, further supporting prominent host extinction despite issues with flux calibration. 

\par

We note a small handful of outliers with $A_V^{host} \lesssim 0.03$ and $c \gtrsim 0.35$. ZTF19aanyuyh (SN~2020pkj; Fig.~\ref{fig:ZTF19aanyuyh_fit}) is the reddest SN~Ia in our primary sample ($c = 0.72\pm0.03$) and we find negligible host extinction (although with large uncertainties) for the VM model, which provides the best overall agreement. Spectra of ZTF19aanyuyh however do show prominent \ion{Na}{i}~D absorption. All other models also predict larger $A_V^{host}$ ($\gtrsim$1~mag). During our fits, we assume an exponential prior on the host $E(B-V)$ with $\lambda = 0.11$ and typically uniform priors on properties related to the explosion. With these priors, ZTF19aanyuyh is best fit by an intrinsically red VM model and the true level of host extinction is under-estimated. Alternative priors, including constraints based on the \ion{Na}{i}~D features, would likely result in a change in the best matching model and may favour an alternative explosion scenario or an intrinsically bluer VM model with increased dust extinction. Indeed, the majority of SNe~Ia in our secondary sample with high $c$ and low host extinction are VM models. This likely results from this scenario being able to produce models with the reddest intrinsic colours. Therefore while these outliers could represent intrinsically red SNe~Ia that do not experience significant host extinction, they may instead be the result of our choice of priors and underestimate $A_V^{host}$. Such outliers are also likely the result of the weaker correlation between $A_V^{host}$ and $c$ found in our sample compared to \cite{sarin--26}.

\subsection{Host galaxy properties}
\label{sect:host}

Figure~\ref{fig:host} gives the distributions of host galaxy properties for each model type. Our results show some indications of a link between the global host galaxy mass and the mass of the exploding white dwarf, with sub-Chandrasekhar mass progenitors exhibiting a slight bias towards higher mass galaxies. The VM model is not favoured for any SNe~Ia hosted by galaxies with masses below $\sim10^{9.3}~M_{\odot}$ in our primary sample. The DDT model however is favoured for the SNe~Ia in both the lowest ($\sim10^{7.5}~M_{\odot}$) and highest ($\sim10^{11}~M_{\odot}$) mass galaxies. As with the distributions of SALT2 parameters however, we caution that this is based on only a small number of SNe~Ia and is not statistically significant. Likewise, our secondary sample shows a similar spread in properties for both the Chandrasekhar and sub-Chandrasekhar mass progenitors (although again these fits may not be robust). 

\par

The mass step is a common approach to correcting for SNe~Ia host galaxy relationships \citep{sullivan--10}. Effectively, SNe~Ia occurring in galaxies above or below the step location are assumed to have different standardised magnitudes and hence an offset is applied either side of the step. Typically this step occurs at $10^{10}~M_{\odot}$, which is close to the lowest galaxy mass for VM models in our primary sample. Indeed, this is mainly driven by a single SN and the rest of our VM models are found in galaxies with masses \textgreater$10^{9.8}~M_{\odot}$. The mass step has been argued to result from differences in the progenitor metallicity or age, which correlate with galaxy mass \citep{tremonti--04}. In terms of the metallicity, the mass step itself could arise from the fact that higher metallicity progenitors are expected to result in lower $^{56}$Ni masses \citep{timmes--03, travaglio--05}. This variation however may not be sufficient to explain the scatter in SNe~Ia luminosities \citep{howell--09}. \cite{sarin--26} report a correlation between $^{56}$Ni mass and host galaxy mass for their fits to the ZTF SN~Ia DR2 sample indicating lower $^{56}$Ni masses are indeed found in higher mass galaxies. Based on our fits, we find a similar (but weak) correlation between $^{56}$Ni and host galaxy mass, but again caution that our $^{56}$Ni mass estimates are unlikely to be fully robust. Nucleosynthesis within each explosion scenario could be affected by differences in progenitor metallicity, inducing some scatter among the resulting SNe, but our results suggest that the mass step may instead be related to the progenitor age. Specifically, the preference for VM models in high mass galaxies would imply that the mass step reflects a real difference in the dominant explosion mechanism and hence progenitor ages of SNe~Ia either side of step.

\par

Similar steps are also reported for other properties of the SN host galaxy or local environment, including the local colour \citep{roman--18}. Indeed, \cite{kelsey--21} report that a local step in $U-R$ colour for the Dark Energy Survey three year cosmology sample \citep{brout--19} is both larger and more significant than a global mass step. In addition, SNe~Ia with bluer local environments appear to represent a more homogeneous sample \citep{kelsey--21}. In Fig.~\ref{fig:host} we show the distributions of local $g-z$ colours for each model type. All explosion scenarios are found across the full colour range and therefore we do not find that the homogeneity of SNe~Ia in bluer environments can be attributed to arising from a specific explosion scenario. In contrast however, SNe~Ia in the reddest environments may predominantly favour a specific type of explosion: the VM scenario. Our results provide some indication that the VM scenario is favoured for SNe~Ia in our primary sample with the reddest local colours, but the differences in our secondary sample are less pronounced and all explosion scenarios are approximately consistent with the same underlying distributions. The local colour is a tracer of recent star formation and if VM models do indeed show a preference for the reddest environments, this could further support the conclusion that only the VM scenario can reproduce the oldest population while the spread in local colours for VM models may reflect the spread in their delay times \citep{rigault--13, rigault--20}.

\par

Based on the star formation rate and host galaxy mass, \cite{ramaiya--25} split their sample of SNe~Ia into three regions: low-mass hosts, high-mass star-forming hosts, and high-mass passive hosts. \cite{ramaiya--25} investigate the properties of SNe~Ia within each region and find that those SNe with high-mass passive hosts show both higher luminosities post-standardisation and lower scatter than other SNe~Ia. Rather than those in locally blue environments, these SNe~Ia may therefore form the most homogeneous sample \citep{ramaiya--25}. Our results would generally support this conclusion and indicate that SNe~Ia in high-mass passive hosts most likely originate from the VM scenario. \cite{ramaiya--25} define high-mass hosts as those with masses \textgreater$10^{10}~M_{\odot}$ and passive as those that do not fall within $3\sigma$ of the main sequence of star-forming galaxies \citep{brinchmann--04, daddi--07, elbaz--07, noeske--07}. For the ZTF SN~Ia DR2 sample we do not have direct measurements of the local star formation rate, therefore we make a simplifying assumption that passive galaxies are those with redder colours than the median ($g-z = 1.08$~mag). Splitting our primary sample into similar regions as those defined by \cite{ramaiya--25}, we find that our high-mass passive region is indeed dominated by VM models, accounting for approximately 60~per~cent of the sample in this region, compared to 25~per~cent overall. As with our previous analysis, this is based on a small number of SNe~Ia -- we find only seven SNe~Ia in our high-mass passive region. Assuming a redder cut-off for passive galaxies of $g-z = 1.36$~mag would produce a pure (but incomplete) sample of only VM models. 

\par

Our analysis is limited by a small number of SNe and few spectra per SN. Larger samples with additional spectra are required to verify the robustness of our results. Nevertheless, our results do provide some hints that environmental steps may be related to real physical differences in the explosion mechanisms and progenitors of SNe~Ia in cosmological samples. Specifically, VM models (i.e. typically older progenitors) appear to be favoured for SNe~Ia in higher mass galaxies and/or with redder local environments. Rather than light curve parameters, which show significant overlap between models, the global and local host galaxy properties may provide the best method of producing homogeneous samples of SNe~Ia arising from a single explosion mechanism.

\par

\subsection{Standardisation}
\label{sect:standard}

Having explored the distributions of SALT2 and host galaxy properties for our explosion models, we now examine how these different samples are affected by SNe~Ia standardisation. We use the \textsc{standax} package presented by \cite{ginolin--25a, ginolin--25b} and follow the standardisation procedure outlined therein. The goal of standardisation is typically to minimise scatter in the so-called Hubble residuals:
\begin{equation}
\label{eqn:hubble_resid}
    \Delta \mu = \mu_o - \mu_c,
\end{equation}
where $\mu_c$ is the distance modulus calculated under a certain cosmological model. Here we assume a flat $\Lambda$CDM cosmology following \cite{planck--20}. The observed distance modulus, $\mu_o$, is determined from the Tripp equation \citep{tripp--98} and given by
\begin{equation}
\label{eqn:tripp}
    \mu_o = m_B - M_0 + \alpha x_1 - \beta c - \gamma p,
\end{equation}
where $m_B$ is the SALT2 $B$-band peak apparent magnitude ($-2.5\log x_0 + 10.5$) and $M_0$ is the standardised peak absolute magnitude. The parameters $\alpha$, $\beta$, and $\gamma$ are global parameters that account for correlations with light curve shape ($x_1$), colour ($c$), and host environmental properties, respectively. Following \cite{ginolin--25a, ginolin--25b}, $p$ is the probability of a given SN falling below the step for a specific environmental tracer (such as global host mass) when measurement uncertainties are taken into account. Using \textsc{standax} to fit our initial cosmology sample of 979 SNe~Ia with local $g-z$ colour as our environmental tracer, we find $\alpha = 0.17\pm0.01$, $\beta = 3.13\pm0.05$, $\gamma = 0.15\pm0.02$, and $M_0 = -19.30\pm0.01$. These values are consistent with those found by \cite{ginolin--25b}. 

\par

\begin{figure*}
\centering
\includegraphics[width=\textwidth]{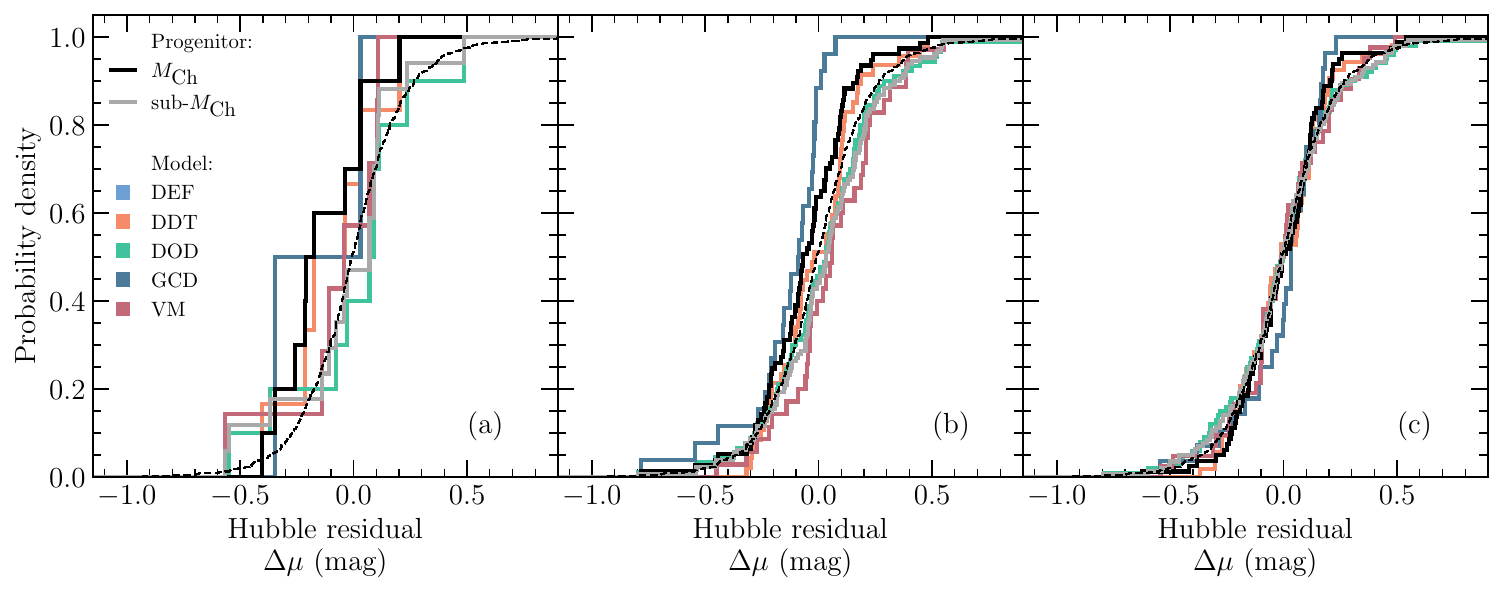}
\caption{Cumulative probability distributions of Hubble residuals assuming no environmental corrections for our cosmology sample (dashed line) and different explosion mechanisms or progenitor scenarios. \textit{Panels a \& b} show Hubble residuals for our primary and secondary samples, respectively, calculated based on parameters determined from fits to the full cosmology sample. \textit{Panel c} shows our combined primary plus secondary sample where Hubble residuals are calculated based on parameters determined from fits to each explosion mechanism or progenitor scenario independently.
}
\label{fig:hubble_residuals}
\centering
\end{figure*}

As evidenced by the hints of changes in galaxy properties discussed in Sect.~\ref{sect:host}, the environmental dependencies applied in Eqn.~\ref{eqn:tripp} may already account for at least some of the scatter arising from different explosion or progenitor scenarios within cosmology samples. We therefore refit our cosmology sample with $\gamma = 0$ and find $\alpha = 0.14\pm0.01$, $\beta = 3.07\pm0.06$, and $M_0 = -19.29\pm0.01$. The resulting Hubble residuals have a mean of 0.00~mag and standard deviation of 0.24~mag, and are shown in Fig.~\ref{fig:hubble_residuals}. Figures~\ref{fig:hubble_residuals}(a) and (b) also show the Hubble residuals for each explosion and progenitor scenario in our primary and secondary samples, respectively, calculated based on this fit to the full cosmology sample. Among the different explosion and progenitor scenarios, the mean Hubble residuals range from $-0.16$ -- 0.07~mag with standard deviations ranging from 0.19 -- 0.29~mag. All means are $\gtrsim$0.01~mag from $\overline{\Delta \mu} = 0.0$. In both samples, Chandrasekhar mass progenitors are skewed towards negative Hubble residuals (i.e. brighter SNe~Ia) and show a mean $0.1$~mag below the mean of sub-Chandrasekhar mass progenitors. Combining our primary and secondary samples (due to the small number of objects), we perform independent fits to each explosion and progenitor scenario and calculate their Hubble residuals accordingly (Fig.~\ref{fig:hubble_residuals}(c)). Unsurprisingly, this removes the offsets in the mean Hubble residuals for each case and all are within $\textless0.01$~mag of $\overline{\Delta \mu} = 0.0$. Our results would therefore suggest that selecting SNe~Ia arising from specific explosion scenarios and performing independent standardisation could lead to reduced scatter in Hubble residuals for future cosmology surveys.

\par

\begin{figure}
\centering
\includegraphics[width=\columnwidth]{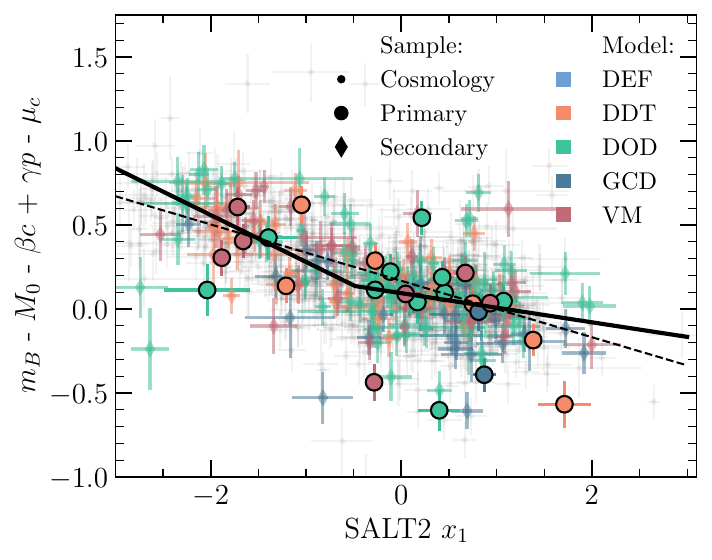}
\caption{Hubble residuals with no $x_1$ correction as a function of $x_1$. The dashed and solid lines show the best-fitting linear and broken $\alpha$ corrections to the Tripp equation. Adapted from fig.~8 of \protect\cite{ginolin--25b}.
}
\label{fig:alpha_residuals}
\centering
\end{figure}

\cite{ginolin--25b} also explore modified corrections in which $\alpha$ is non-linear. Specifically, $\alpha$ in Eqn.~\ref{eqn:tripp} is replaced by a piecewise function $\mathcal{A}(x_1)$ equal to $\alpha_0$ for $x_1$ $\textless$ $x_1^0$ and $\alpha_1$ for $x_1$ $\geq$ $x_1^0$. \cite{ginolin--25b} argue that this model provides a significantly better fit to the observations than the typical linear model assumed in cosmological analyses. Applying this model we find $\alpha_0 = 0.27\pm0.01$, $\alpha_1 = 0.09\pm0.01$, and $x_1^0 = -0.48\pm0.08$, which are consistent with the values reported by \cite{ginolin--25b}. In addition, we find significant differences in $\beta$, $\gamma$, and $M_0$ compared to a linear model. Assuming a non-linear $\alpha$ results in an increased colour dependence ($\beta = 3.29\pm0.03$), larger local colour step ($\gamma = 0.18\pm0.01$), and brighter standardised magnitude ($M_0 = -19.44\pm0.02$). In Fig.~\ref{fig:alpha_residuals} we show standardised Hubble residuals ($\alpha = 0$) for our cosmology sample and how this relates to each explosion scenario (see fig.~8 of \citealt{ginolin--25b}). As shown by Fig.~\ref{fig:alpha_residuals}, all explosion scenarios are found on either side of the $x_1$ break location, but different scenarios may dominate different sides. In both our primary and secondary samples, the majority of DOD models are found with $x_1 \textgreater x_1^0$. Conversely $x_1 \textless x_1^0$ seems to show a larger relative fraction of VM models. On the other hand, DDT models appear relatively uniform across $x_1$. Therefore the non-linear $\alpha$ found by \cite{ginolin--25b} may be due to changes in the relative populations of explosion models and their need to be standardised independently. 

\par

\begin{figure}
\centering
\includegraphics[width=\columnwidth]{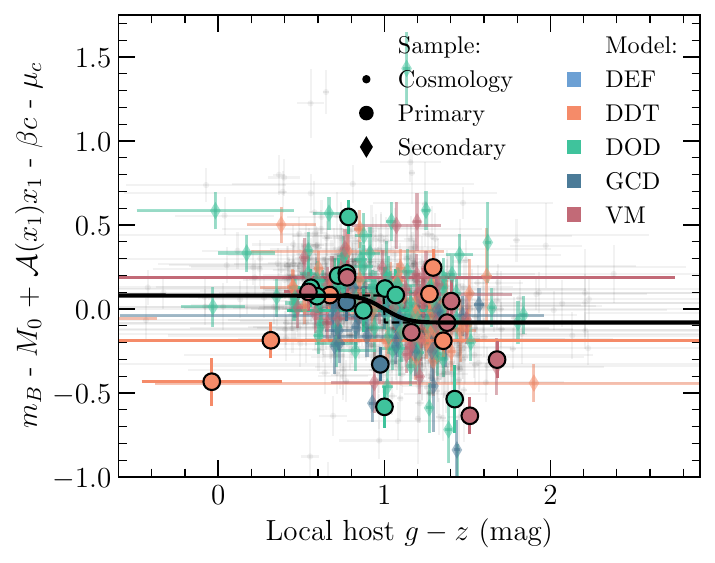}
\caption{Hubble residuals with no environmental correction as a function of local $g-z$ colour. The dashed line shows the step correction applied above and below $g-z = 1.0$~mag. The solid line shows the step correction convolved with a Gaussian ($\sigma = 0.14$~mag). Adapted from fig.~9 of \protect\cite{ginolin--25b}.
}
\label{fig:gamma_residuals}
\centering
\end{figure}

The impact of the local colour step (assuming a non-linear $\alpha$ model) on different explosion scenarios is shown in Fig.~\ref{fig:gamma_residuals} (similar to fig.~9 of \citealt{ginolin--25b}). Here we find that the presence of a step could also be related to changes in the relative populations of explosion models either side of the step. Similar to $x_1$, our primary and secondary samples show that all explosion scenarios are found either side of the step location. As discussed in Sect.~\ref{sect:host} however, VM models appear to be preferentially found in the reddest environments. Figure~\ref{fig:gamma_residuals} therefore shows that these models dominate environments redder than the local colour step. Bluer than the step, we find fewer VM models and an increase in other explosion models. 

\par

Our analysis is limited by systematic biases in our sample, a relatively small number of SNe~Ia, and few spectroscopic observations per SN. We are therefore unable to perform robust statistical analysis of our conclusions. Nevertheless our results hint that at least some scatter in Hubble residuals could be due to the presence of different explosion or progenitor scenarios within cosmological samples. Although our analysis cannot determine the `correct' explosion scenario for each SN, the fact that different scenarios are favoured for different SNe indicates that there are real, intrinsic differences between these SNe and these differences may be due to how the progenitor white dwarf exploded. Selecting pure samples containing only a single explosion or progenitor scenario and standardising independently could be a viable method towards reducing scatter in Hubble residuals and improving the precision of cosmological parameters in future surveys.

\section{Conclusions}
\label{sect:conclusions}
The physical origin(s) of the standardisation corrections used in SNe~Ia cosmology have been studied extensively, but remain poorly-understood. In this work, we demonstrated a new approach to understanding these effects. We presented the first attempt at systematic spectroscopic modelling of a large number of cosmological SNe~Ia and exploring the impact of different explosion and progenitor scenarios on their standardisation.

\par

Beginning from the ZTF SN~Ia DR2 \citep{rigault--25}, we selected a volume-limited sample of SNe~Ia that pass typical cosmology cuts, and meet our signal-to-noise ratio and spectral resolution requirements. We split the resulting objects into a primary sample of 28 SNe~Ia that have at least two spectra and a secondary sample of 216 SNe~Ia with only one spectrum. We then fit all spectra using \textsc{riddler} \citep{magee--24}, a machine learning framework for automated and quantitative fitting of SNe~Ia spectral sequences trained on predictions from realistic explosion simulations. In a companion paper, \cite{magee--26a} present recent updates to \textsc{riddler}, such as more robust uncertainty estimates and a broader range of explosion scenarios. This includes pure deflagrations (DEF; \citealt{fink-2014}), delayed detonations (DDT; \citealt{seitenzahl--13}), double detonations (DOD; \citealt{gronow--21}), gravitationally confined detonations (GCD; \citealt{lach--22b}), and violent mergers (VM; \citealt{pakmor--10, pakmor-2012, kromer--13b}). Using \textsc{riddler}, we determined the best matching explosion scenario for each SN in our sample. We presented representative fits and showed that they are generally able to reproduce the main spectroscopic features observed. Across both our primary and secondary samples, we found that approximately two thirds of SNe~Ia are best reproduced by explosions of sub-Chandrasekhar mass white dwarfs. In particular, our results indicated a slight preference for the DOD scenario over VM. Among those SNe~Ia favouring explosions of Chandrasekhar mass white dwarfs, we found a preference for the DDT scenario over both the DEF and GCD scenarios. For a small number of cases, our fits were unable to distinguish between different progenitor or explosion scenarios.

\par

As the first attempt at such an approach, our analysis highlighted a number of challenges and limitations associated with large-scale spectroscopic modelling of SNe~Ia and how this may be applied to future cosmology surveys. \textsc{riddler} is designed primarily for fitting spectroscopic sequences, as the time evolution of spectral features provides tighter constraints on the explosion mechanism. Despite representing the largest single release of spectroscopic SNe~Ia to date, only $\sim$7~per~cent of the ZTF SN~Ia DR2 sample passed our selection and quality cuts. Furthermore, \textless1~per~cent have multiple spectra. Our primary sample is therefore limited by the inclusion of only a small number of SNe~Ia, while fits to our secondary sample may not be robust due to containing only a single spectrum per SN. Ideally, observations of each SN would cover the pre-maximum, maximum, and post-maximum phases and therefore each SN would have at least three spectra at similar phases. In addition, although the ZTF SN~Ia DR2 is expected to be spectroscopically complete for normal SNe~Ia, the selection of which SNe will be observed by different instruments (and therefore have different spectral resolutions) may be arbitrary. This will affect which SNe pass our cuts and introduce a bias into our samples. Hence the samples that we fit with \textsc{riddler} may not reflect the true distributions of explosion mechanisms or progenitor scenarios found in cosmology samples or the Universe. Another limitation of our approach is that \textsc{riddler} does not determine the `correct' explosion model for a given SN or whether the preferred model represents a `good' fit. Instead it simply ranks the explosion scenarios in the training data based on the level of agreement. Other scenarios not considered may produce better agreement.

\par

The limitations in our sample selection and size mean that we were unable to asses the statistical significance of any trends found within our sample. Nevertheless, the fact that we do not find all SNe~Ia are best reproduced by the same explosion scenario demonstrates that there are real, intrinsic differences between their spectra and these differences may be related to their explosion physics. Our results are therefore useful for demonstrating the types of explosion scenarios that could plausibly be found in cosmology samples and their potential links with standardisation corrections. With these limitations in mind, our results showed some indications of differences in the SALT2 $x_1$ parameters \citep{guy--07} among different models. We found that explosions of Chandrasekhar mass white dwarfs are not favoured for the fastest evolving SNe~Ia, while explosions of sub-Chandrasekhar mass white dwarfs are spread relatively evenly across $x_1$. Likewise, sub-Chandrasekhar mass explosions appear to be favoured for the reddest SNe~Ia, but are also found across a range of colours. Considering the host galaxy properties, our results indicated that the VM scenario is biased towards higher mass galaxies and redder local environments. This suggests that selecting SNe~Ia in massive, passive galaxies could be a viable strategy for producing a homogeneous sample arising from the same explosion mechanism. 

\par

Finally, we also explored how each of the different explosion and progenitor scenarios is affected by standardisation. Without considering any environmental corrections, we showed that applying a single light curve shape and colour correction ($\alpha$ and $\beta$, respectively) to the entire ZTF SN~Ia DR2 sample leads to a bias in which Chandrasekhar mass explosions typically have more negative Hubble residuals (i.e. brighter SNe~Ia) than sub-Chandrasekhar mass explosions. Fitting and standardising each explosion scenario with their own independent $\alpha$ and $\beta$ could lead to improved corrections and reduced scatter. We investigated claims of a non-linear $\alpha$ correction and a magnitude offset depending on the local host $g-z$ colour \citep{ginolin--25a, ginolin--25b}. We showed that both phenomena could potentially arise from changes in the relative populations of explosion models. Specifically, low $x_1$ and high $g-z$ appear to show a larger fraction of VM models that approximately coincides with the locations of the $\alpha$ break and local colour step.

\par

The methodology outlined in this work can be readily applied to other samples and may be the key to uncovering the SN physics driving standardisation of cosmological SNe~Ia. Indeed, SNe~Ia physics represents the largest source of systematic uncertainty in current cosmological analyses and reducing this uncertainty will be critical for improving constraints of cosmological parameters \citep{vincenzi--24}. To achieve this however, our results highlight that a change in follow up strategy may be necessary. How many SNe~Ia are truly required for cosmology? For current and future surveys, the goal has typically been to observe as many SNe~Ia as possible (thousands or more) to reduce statistical uncertainties, but systematic uncertainties still remain. We have shown that selecting clean samples of different explosion or progenitor scenarios can reduce systematic uncertainties. Light curve properties alone are insufficient for producing such clean samples, hence the utility of large samples of SNe~Ia with few or no spectroscopic observations will remain somewhat limited. Alternatively, observing fewer SNe and more spectra per SN could provide a better avenue for reducing systematics associated with multiple populations. Even observing half as many SNe~Ia, but doubling the spectroscopic observations per SN, would lead to a significant increase over the sample size fit here. With physically motivated modelling approaches \citep{magee--24, sarin--26}, it is now possible to approach such samples from a physical, rather than empirical, basis and uncover the physics driving SN~Ia cosmology.

%

\section*{Acknowledgements}

We thank Conor Byrne, Madeleine Ginolin, Lisa Kelsey, Tom Killestein, Joe Lyman, Miika Pursiainen, and Danny Steeghs for useful discussion. MRM acknowledges a Warwick Astrophysics prize post-doctoral fellowship made possible thanks to a generous philanthropic donation. Computing facilities were provided by the Scientific Computing Research Technology Platform of the University of Warwick and the Queen's University Belfast HPC Kelvin cluster. This research made use of \textsc{Tardis}, a community-developed software package for spectral synthesis in supernovae \citep{tardis}. The development of \textsc{Tardis} received support from the Google Summer of Code initiative and from ESA's Summer of Code in Space program. \textsc{Tardis} makes extensive use of Astropy and PyNE. This work made use of the Heidelberg Supernova Model Archive (HESMA), https://hesma.h-its.org. We derive posterior probability distributions and the Bayesian
evidence with the nested sampling Monte Carlo algorithm
MLFriends (Buchner, 2014; 2019) using the
UltraNest\footnote{\url{https://johannesbuchner.github.io/UltraNest/}} package (Buchner 2021).

\section*{Data Availability}

\texttt{riddler} is publicly available on GitHub\footnote{\href{https://github.com/MarkMageeAstro/riddler}{https://github.com/MarkMageeAstro/riddler}}.



\bibliographystyle{mnras}
\bibliography{mnras_template}




\appendix

\clearpage
\section{Primary sample figures}
\label{sect:primary_figures}

\begin{figure}
    \centering
    \includegraphics[width=\columnwidth]{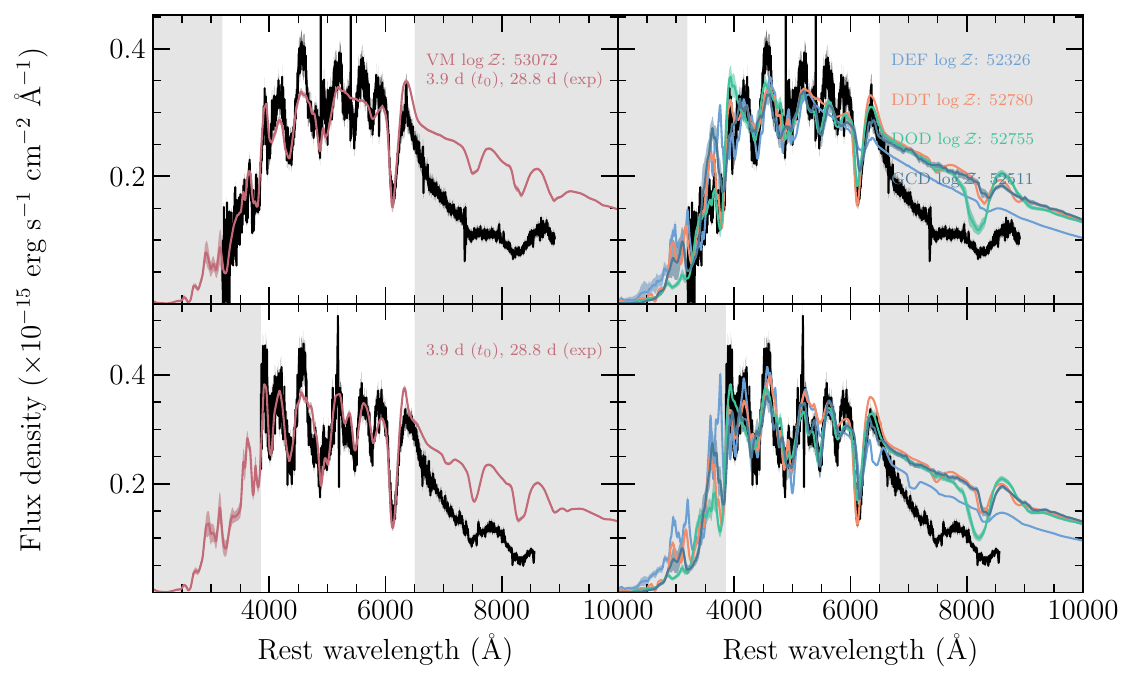}
    \caption{Fits to ZTF18aagtcxj as in Fig.~\ref{fig:ZTF18aahheaj_fit}}
    \label{fig:ZTF18aagtcxj_fit}
\end{figure}

\begin{figure}
    \centering
    \includegraphics[width=\columnwidth]{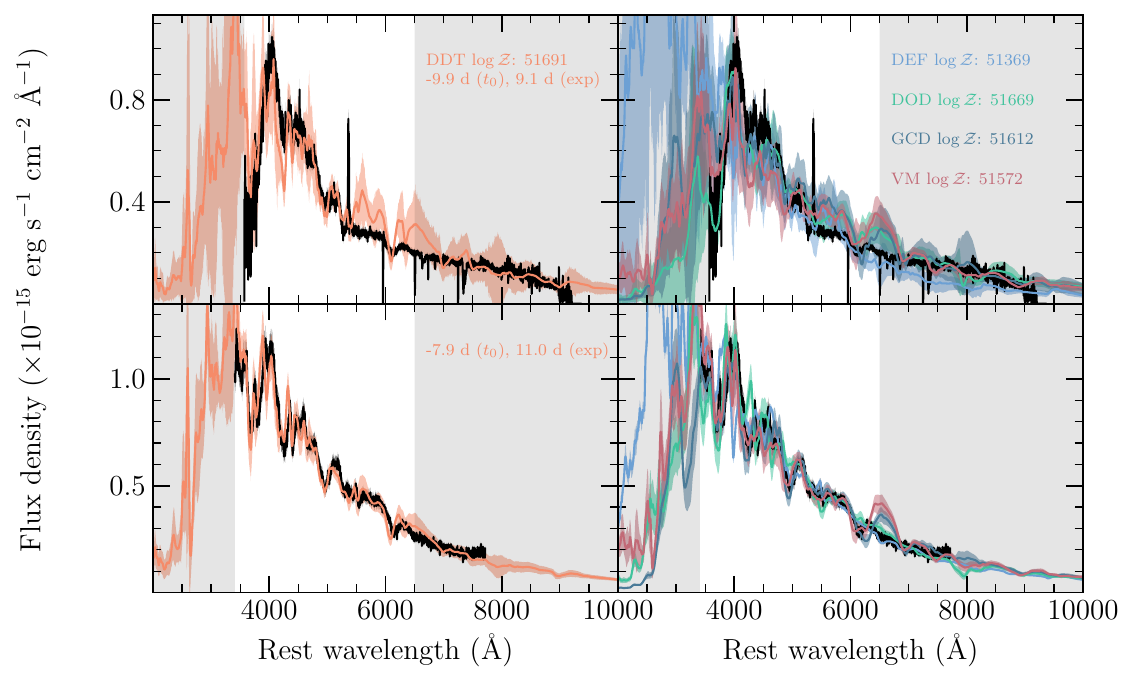}
    \caption{Fits to ZTF18aaumeys as in Fig.~\ref{fig:ZTF18aahheaj_fit}}
    \label{fig:ZTF18aaumeys_fit}
\end{figure}

\begin{figure}
    \centering
    \includegraphics[width=\columnwidth]{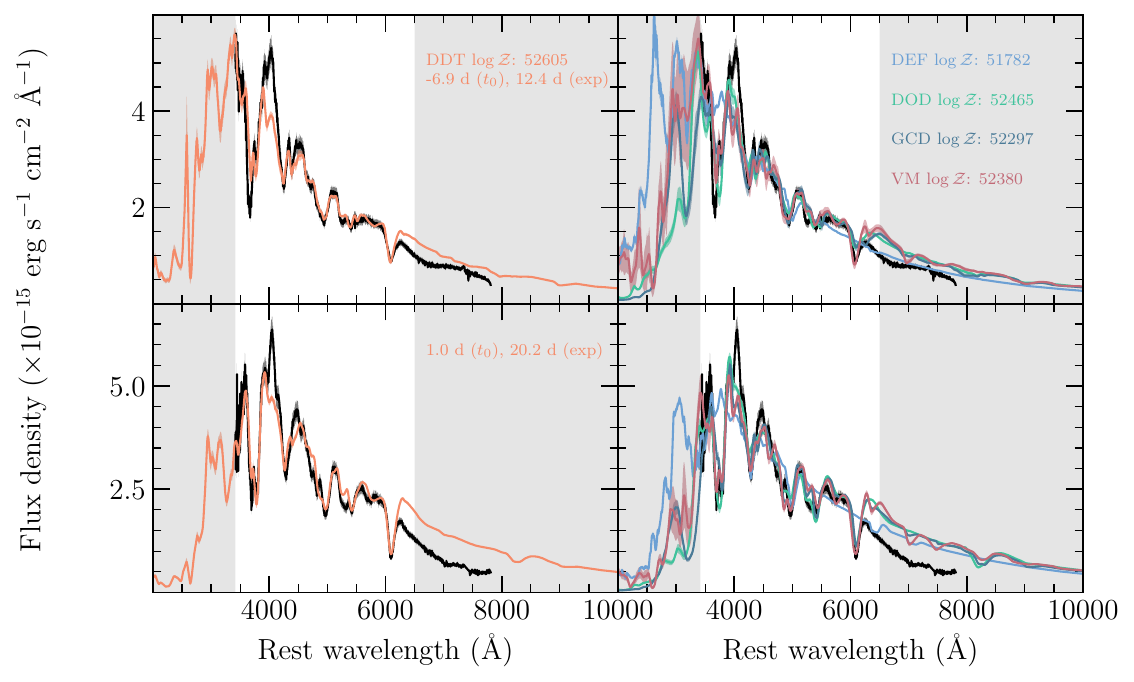}
    \caption{Fits to ZTF18abauprj as in Fig.~\ref{fig:ZTF18aahheaj_fit}}
    \label{fig:ZTF18abauprj_fit}
\end{figure}

\begin{figure}
    \centering
    \includegraphics[width=\columnwidth]{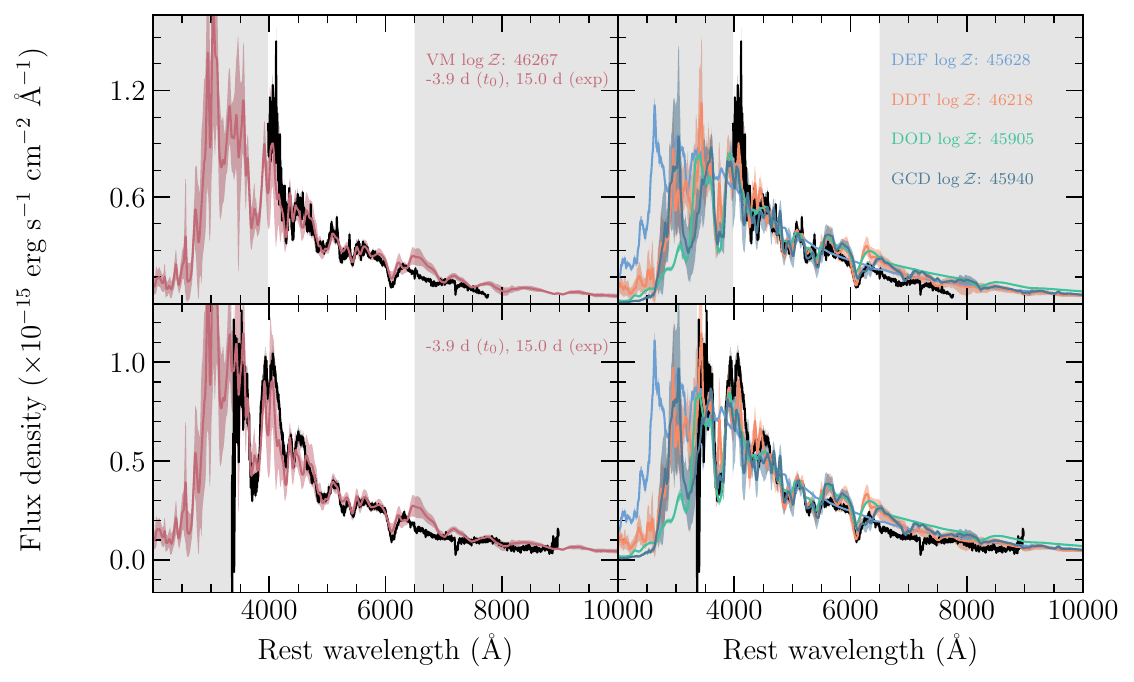}
    \caption{Fits to ZTF18abosdwf as in Fig.~\ref{fig:ZTF18aahheaj_fit}}
    \label{fig:ZTF18abosdwf_fit}
\end{figure}

\begin{figure}
    \centering
    \includegraphics[width=\columnwidth]{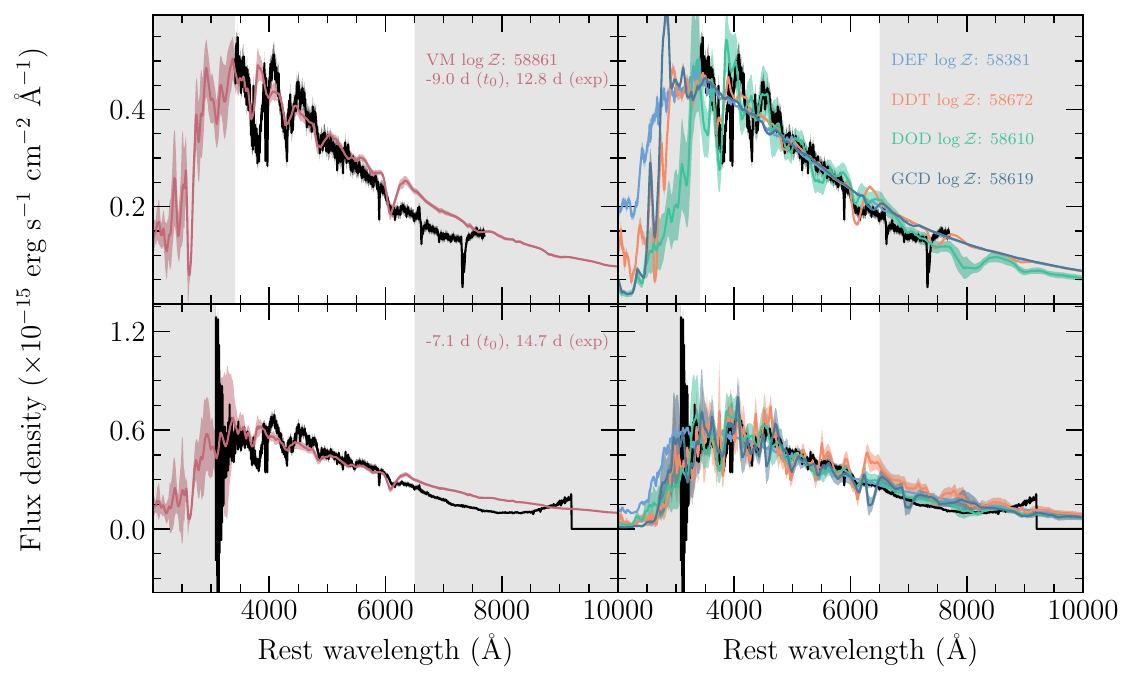}
    \caption{Fits to ZTF18abuhzfc as in Fig.~\ref{fig:ZTF18aahheaj_fit}}
    \label{fig:ZTF18abuhzfc_fit}
\end{figure}

\begin{figure}
    \centering
    \includegraphics[width=\columnwidth]{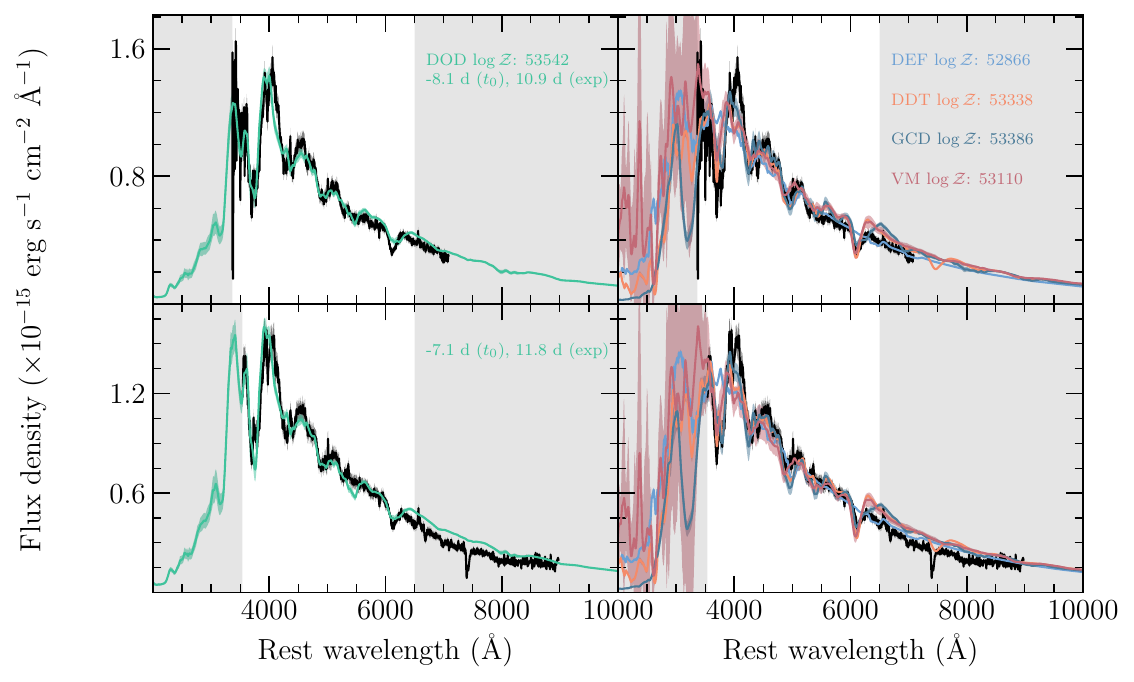}
    \caption{Fits to ZTF18acurmpd as in Fig.~\ref{fig:ZTF18aahheaj_fit}}
    \label{fig:ZTF18acurmpd_fit}
\end{figure}

\begin{figure}
    \centering
    \includegraphics[width=\columnwidth]{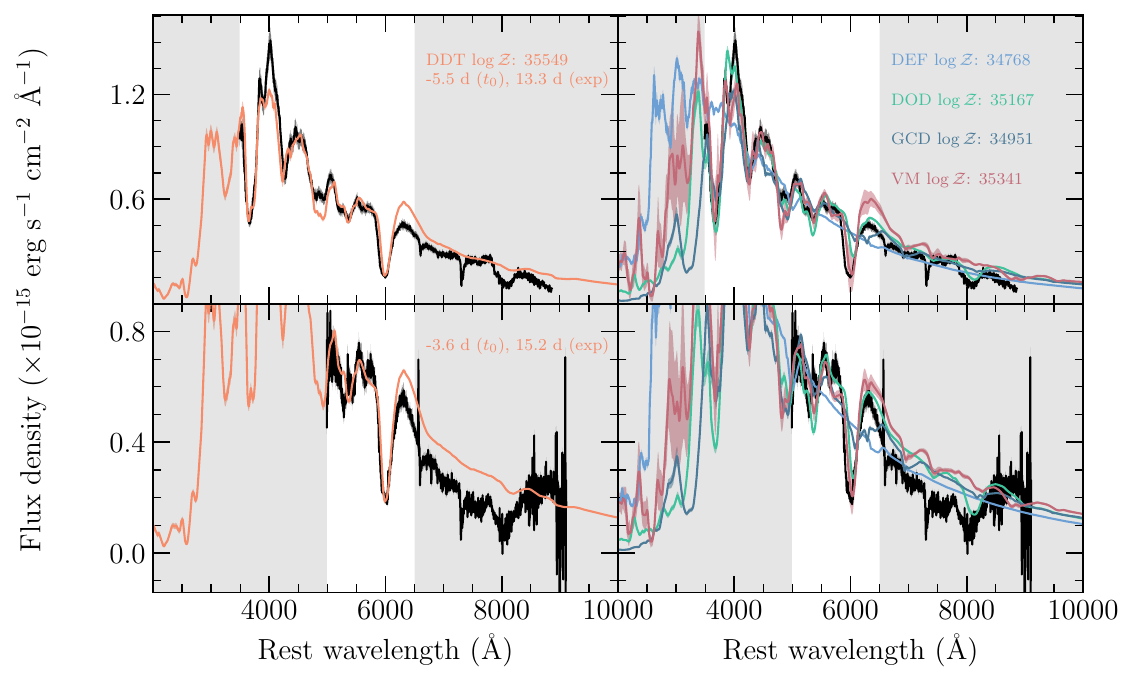}
    \caption{Fits to ZTF18adaifep as in Fig.~\ref{fig:ZTF18aahheaj_fit}}
    \label{fig:ZTF18adaifep_fit}
\end{figure}

\begin{figure}
    \centering
    \includegraphics[width=\columnwidth]{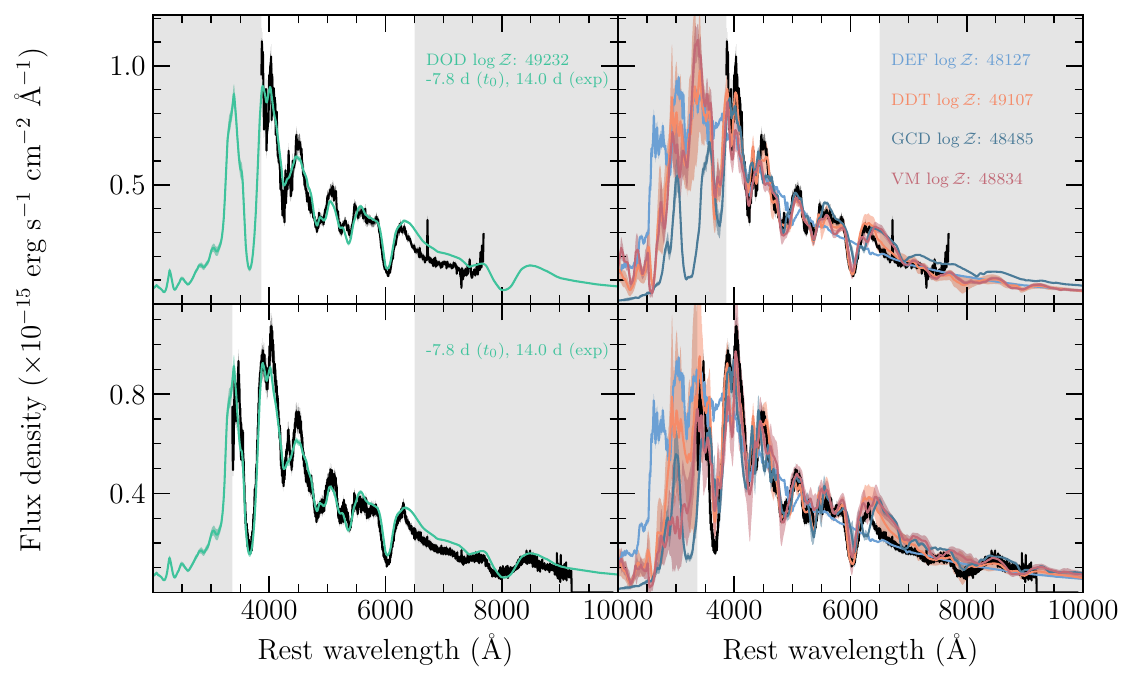}
    \caption{Fits to ZTF19aabmybj as in Fig.~\ref{fig:ZTF18aahheaj_fit}}
    \label{fig:ZTF19aabmybj_fit}
\end{figure}

\begin{figure}
    \centering
    \includegraphics[width=\columnwidth]{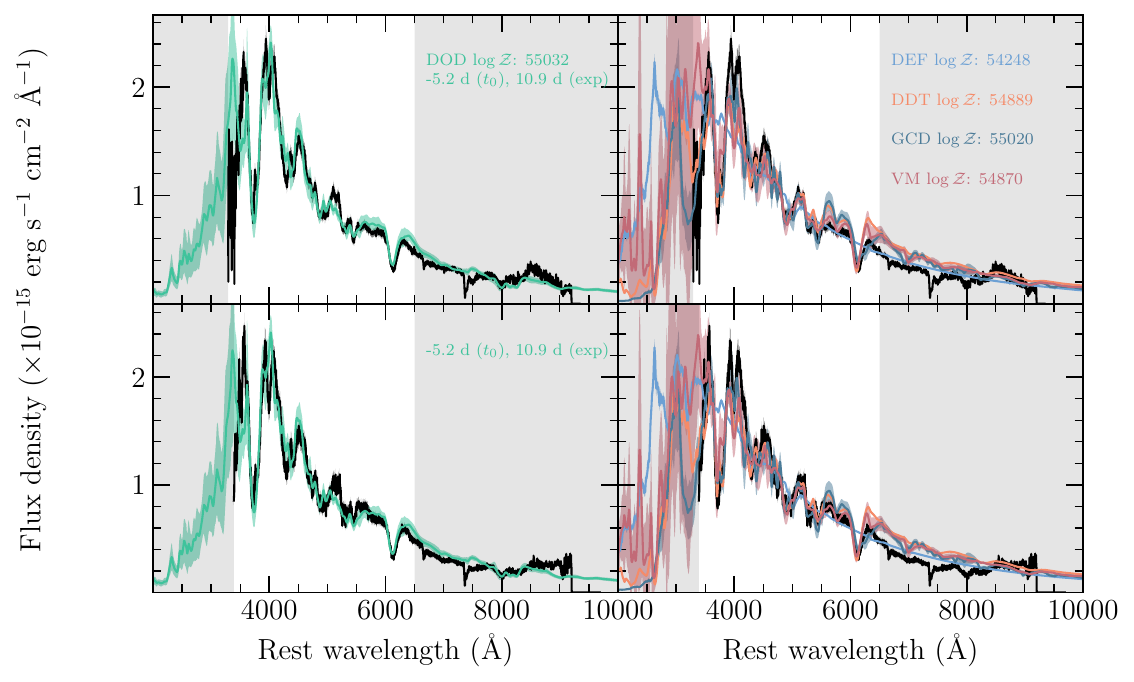}
    \caption{Fits to ZTF19aadnxat as in Fig.~\ref{fig:ZTF18aahheaj_fit}}
    \label{fig:ZTF19aadnxat_fit}
\end{figure}

\begin{figure}
    \centering
    \includegraphics[width=\columnwidth]{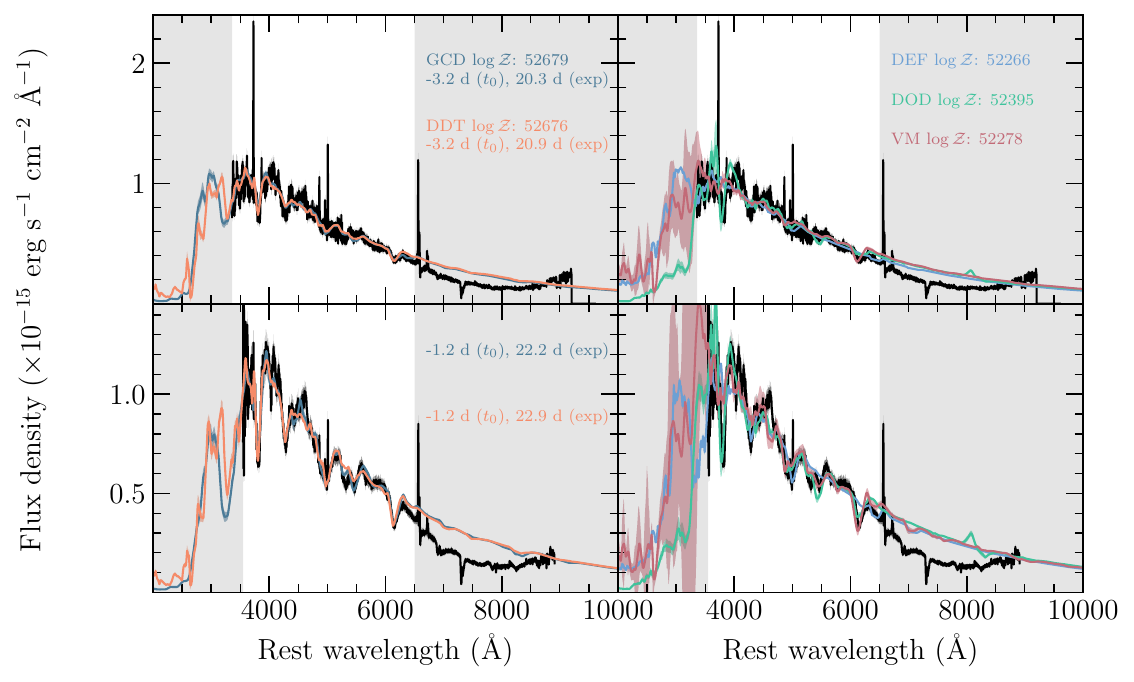}
    \caption{Fits to ZTF19aadoocn as in Fig.~\ref{fig:ZTF18aahheaj_fit}}
    \label{fig:ZTF19aadoocn_fit}
\end{figure}

\begin{figure}
    \centering
    \includegraphics[width=\columnwidth]{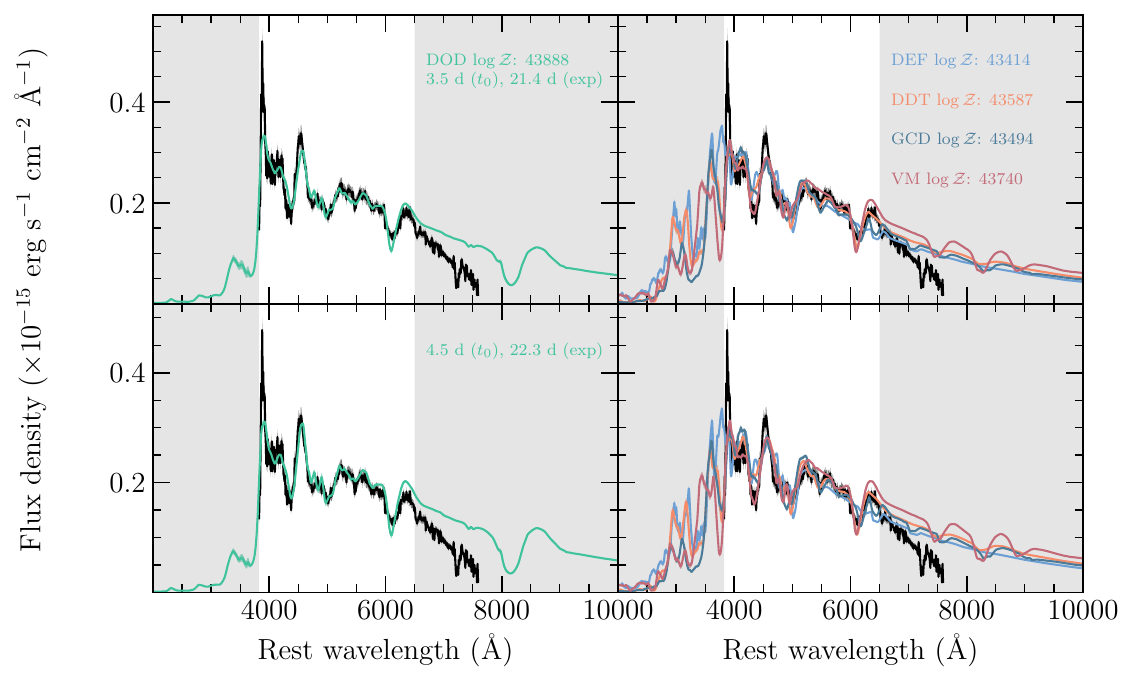}
    \caption{Fits to ZTF19aagkfvq as in Fig.~\ref{fig:ZTF18aahheaj_fit}}
    \label{fig:ZTF19aagkfvq_fit}
\end{figure}

\begin{figure}
    \centering
    \includegraphics[width=\columnwidth]{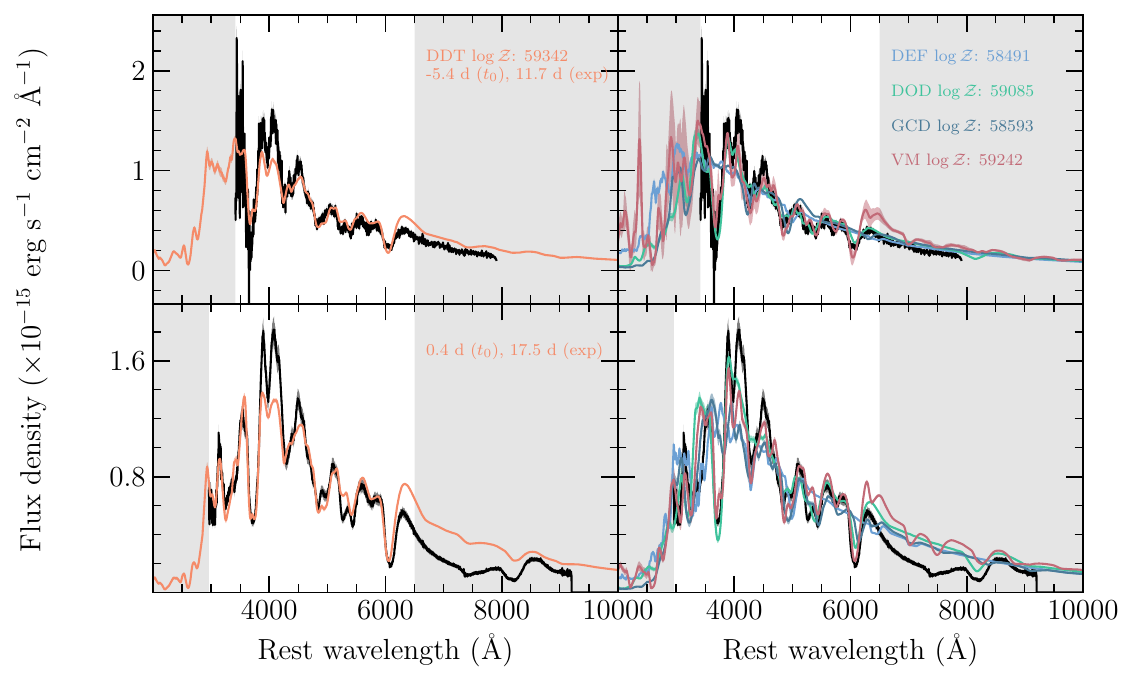}
    \caption{Fits to ZTF19aakoasy as in Fig.~\ref{fig:ZTF18aahheaj_fit}}
    \label{fig:ZTF19aakoasy_fit}
\end{figure}

\begin{figure}
    \centering
    \includegraphics[width=\columnwidth]{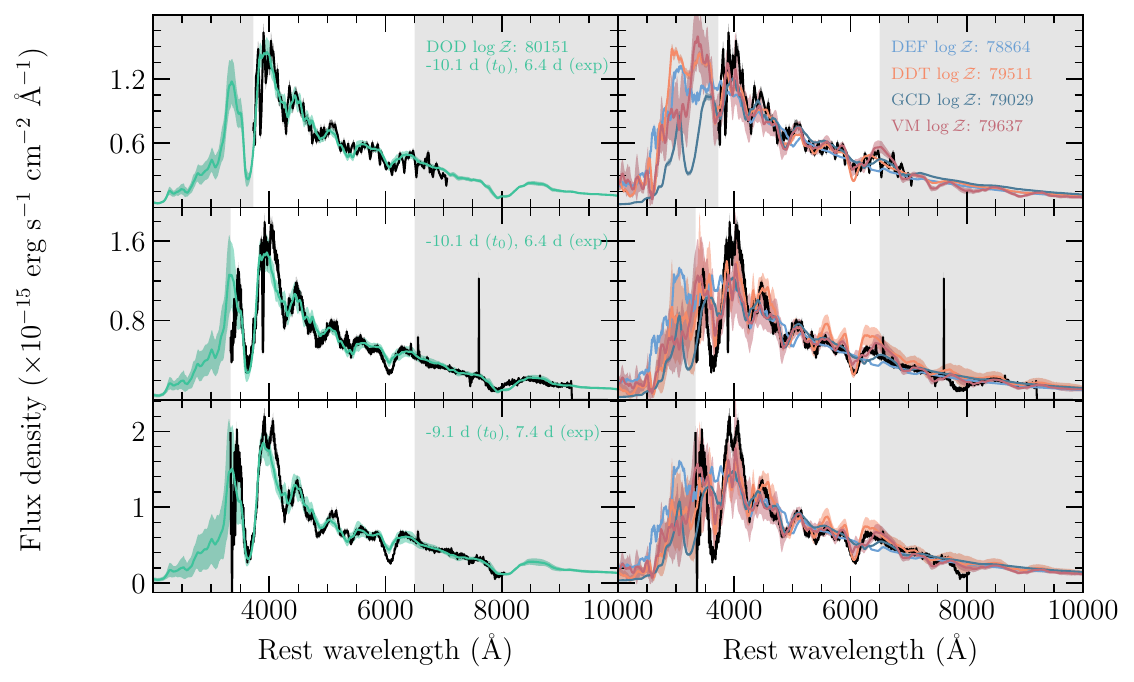}
    \caption{Fits to ZTF19aakzwao as in Fig.~\ref{fig:ZTF18aahheaj_fit}}
    \label{fig:ZTF19aakzwao_fit}
\end{figure}

\begin{figure}
    \centering
    \includegraphics[width=\columnwidth]{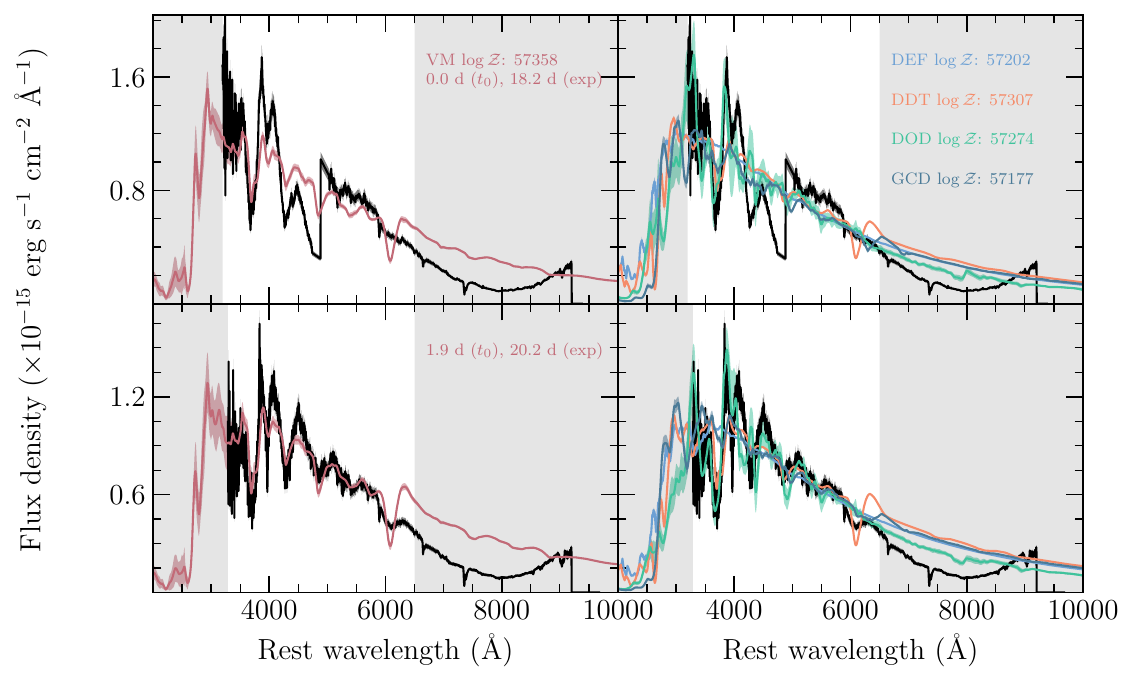}
    \caption{Fits to ZTF19aamkdtq as in Fig.~\ref{fig:ZTF18aahheaj_fit}}
    \label{fig:ZTF19aamkdtq_fit}
\end{figure}

\begin{figure}
    \centering
    \includegraphics[width=\columnwidth]{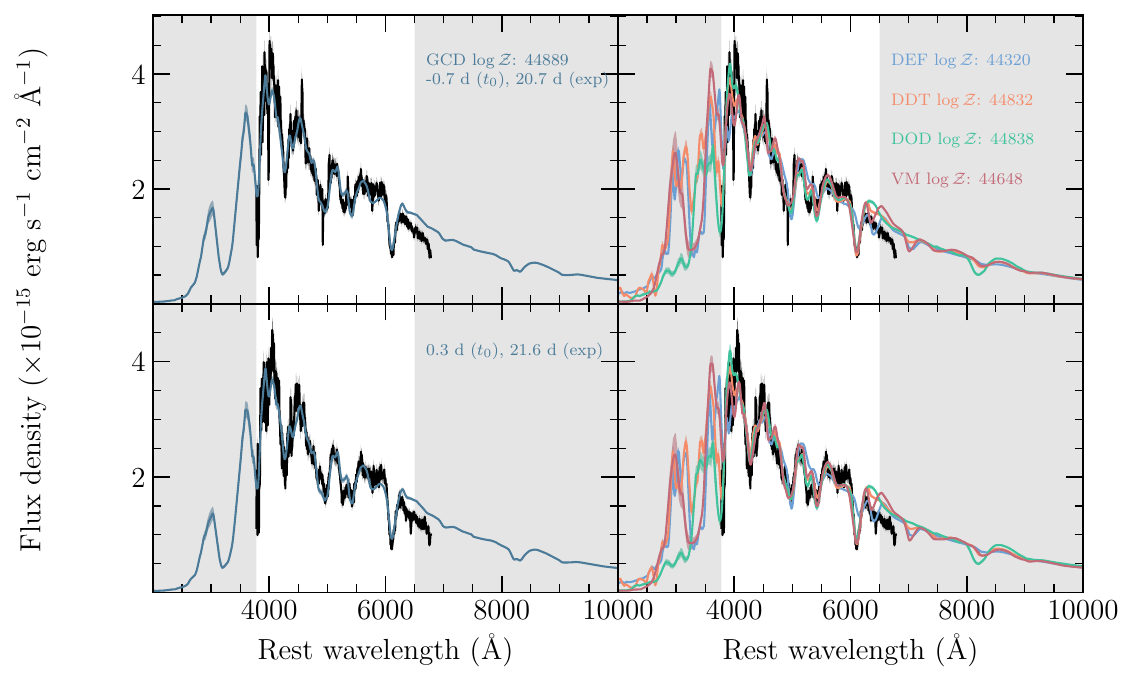}
    \caption{Fits to ZTF19aampqcq as in Fig.~\ref{fig:ZTF18aahheaj_fit}}
    \label{fig:ZTF19aampqcq_fit}
\end{figure}

\begin{figure}
    \centering
    \includegraphics[width=\columnwidth]{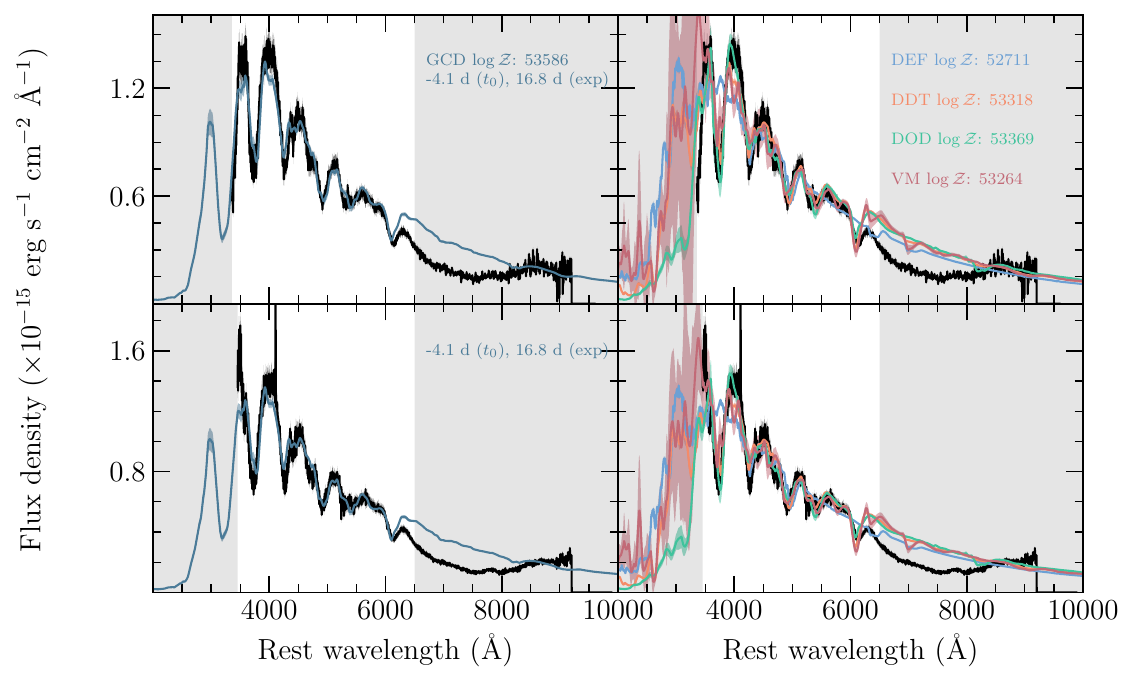}
    \caption{Fits to ZTF19aamrdmm as in Fig.~\ref{fig:ZTF18aahheaj_fit}}
    \label{fig:ZTF19aamrdmm_fit}
\end{figure}

\begin{figure}
    \centering
    \includegraphics[width=\columnwidth]{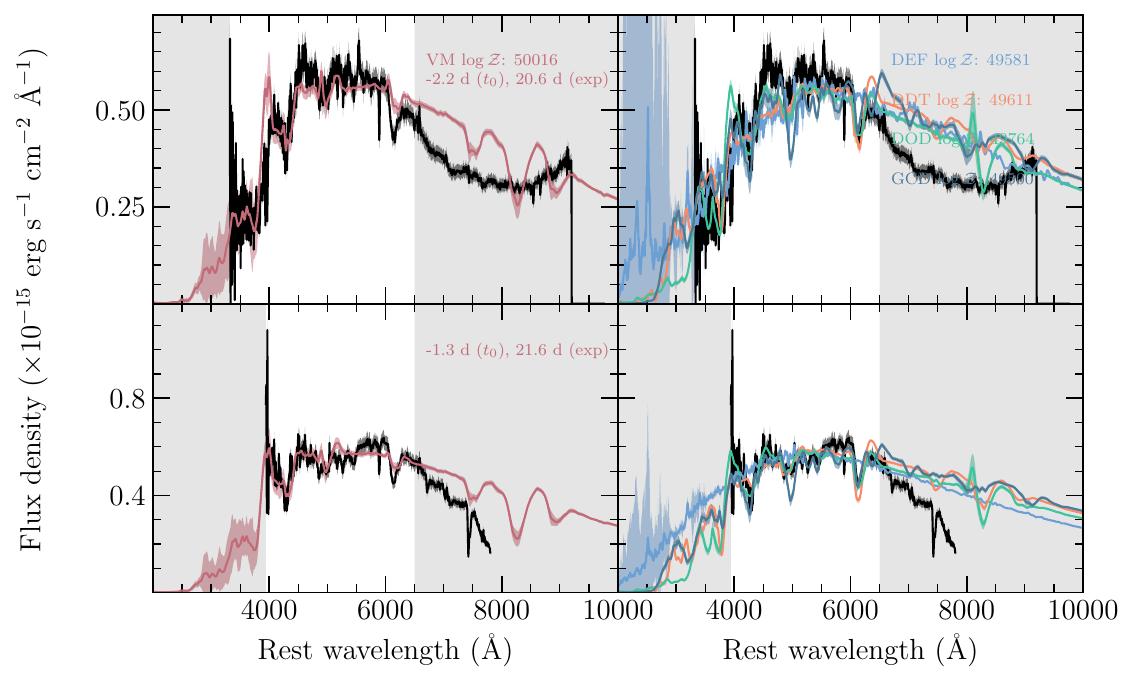}
    \caption{Fits to ZTF19aanyuyh as in Fig.~\ref{fig:ZTF18aahheaj_fit}}
    \label{fig:ZTF19aanyuyh_fit}
\end{figure}

\begin{figure}
    \centering
    \includegraphics[width=\columnwidth]{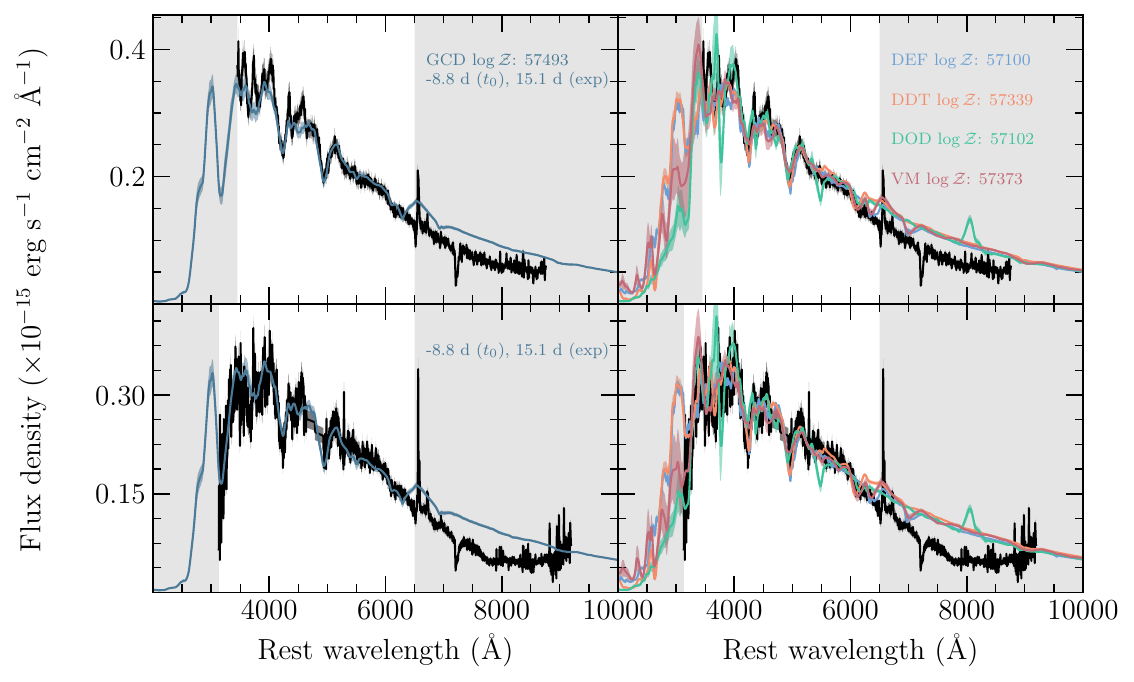}
    \caption{Fits to ZTF19aaxffum as in Fig.~\ref{fig:ZTF18aahheaj_fit}}
    \label{fig:ZTF19aaxffum_fit}
\end{figure}

\begin{figure}
    \centering
    \includegraphics[width=\columnwidth]{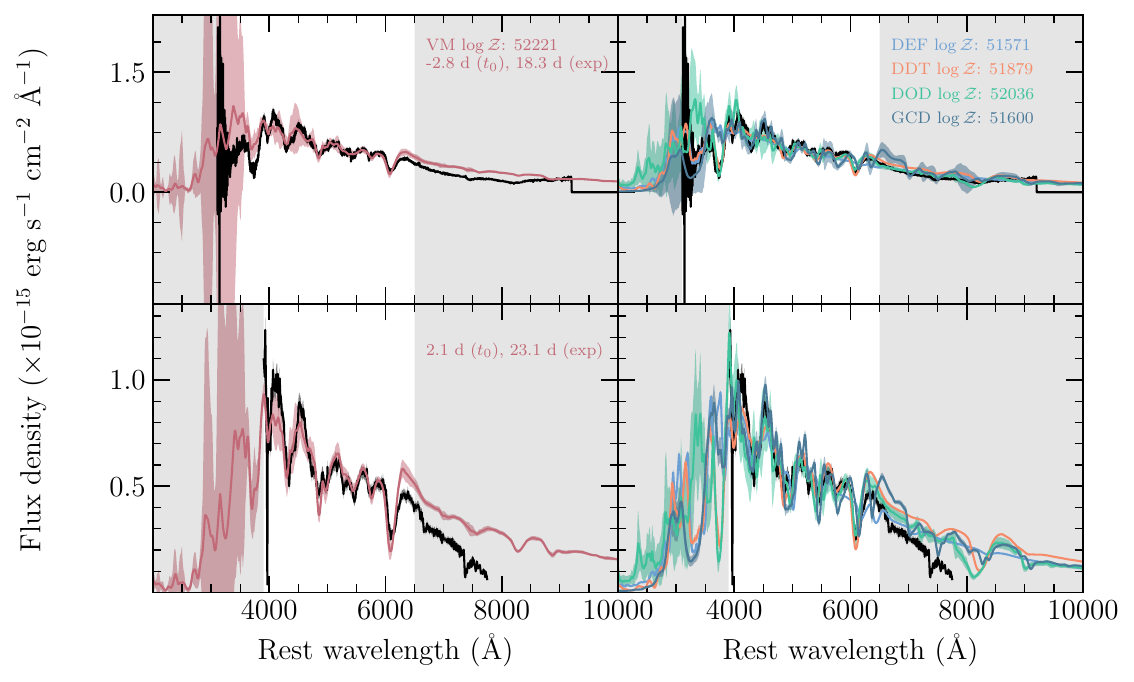}
    \caption{Fits to ZTF19acazbqm as in Fig.~\ref{fig:ZTF18aahheaj_fit}}
    \label{fig:ZTF19acazbqm_fit}
\end{figure}

\begin{figure}
    \centering
    \includegraphics[width=\columnwidth]{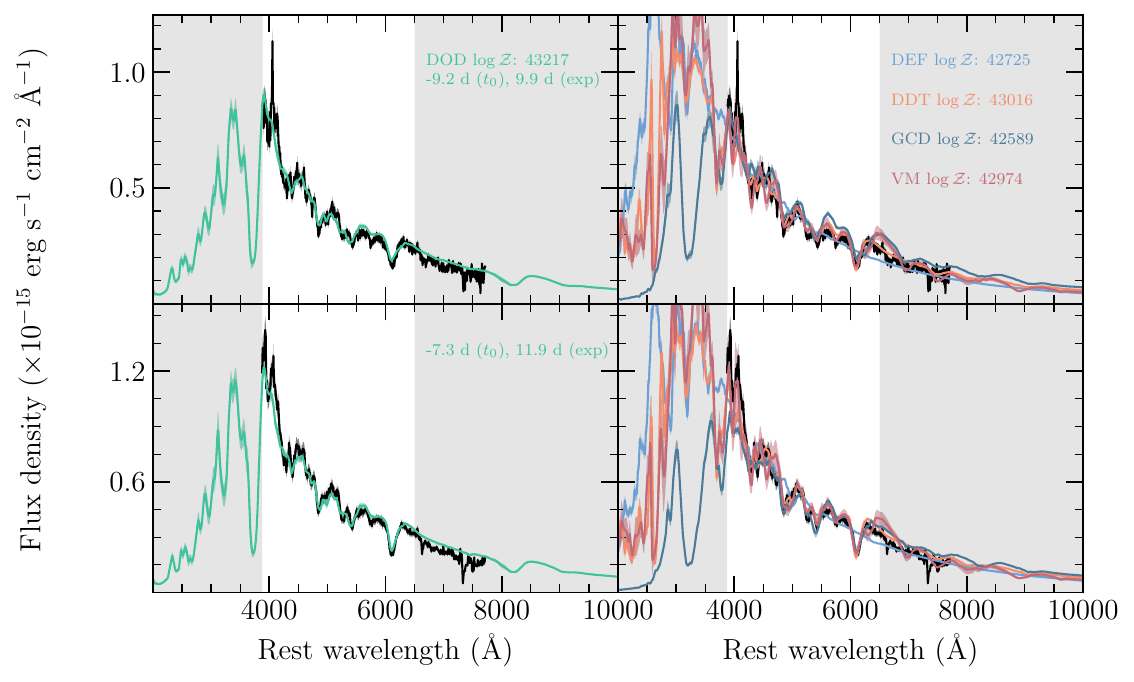}
    \caption{Fits to ZTF19acxyumq as in Fig.~\ref{fig:ZTF18aahheaj_fit}}
    \label{fig:ZTF19acxyumq_fit}
\end{figure}

\begin{figure}
    \centering
    \includegraphics[width=\columnwidth]{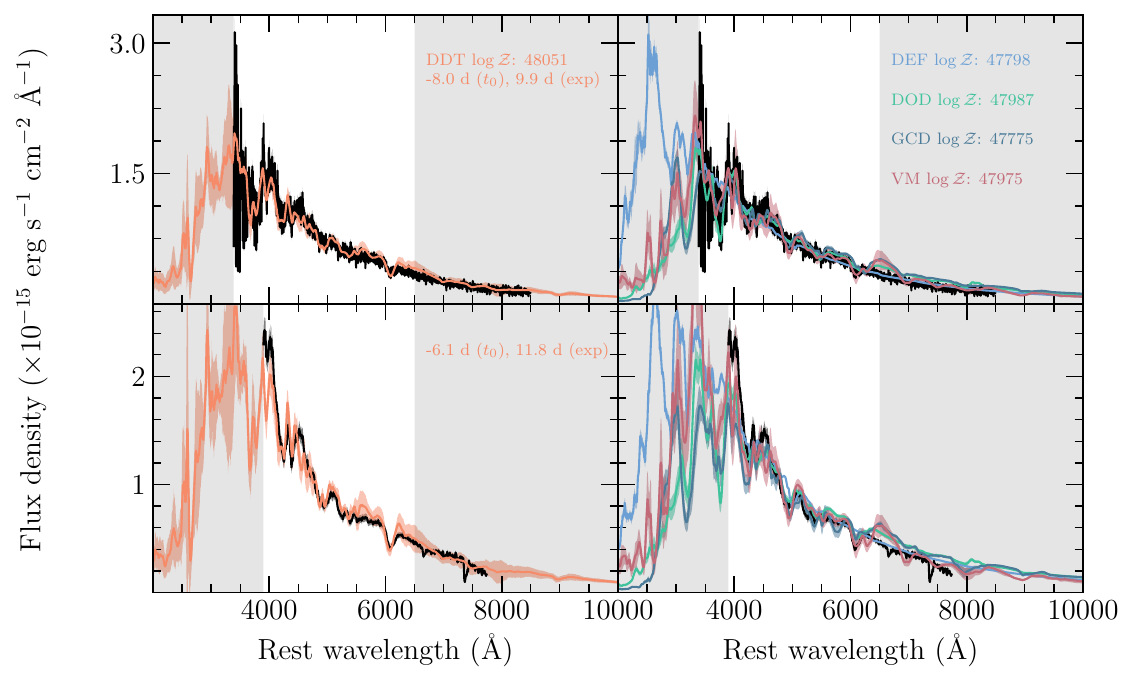}
    \caption{Fits to ZTF19acyiqnw as in Fig.~\ref{fig:ZTF18aahheaj_fit}}
    \label{fig:ZTF19acyiqnw_fit}
\end{figure}

\begin{figure}
    \centering
    \includegraphics[width=\columnwidth]{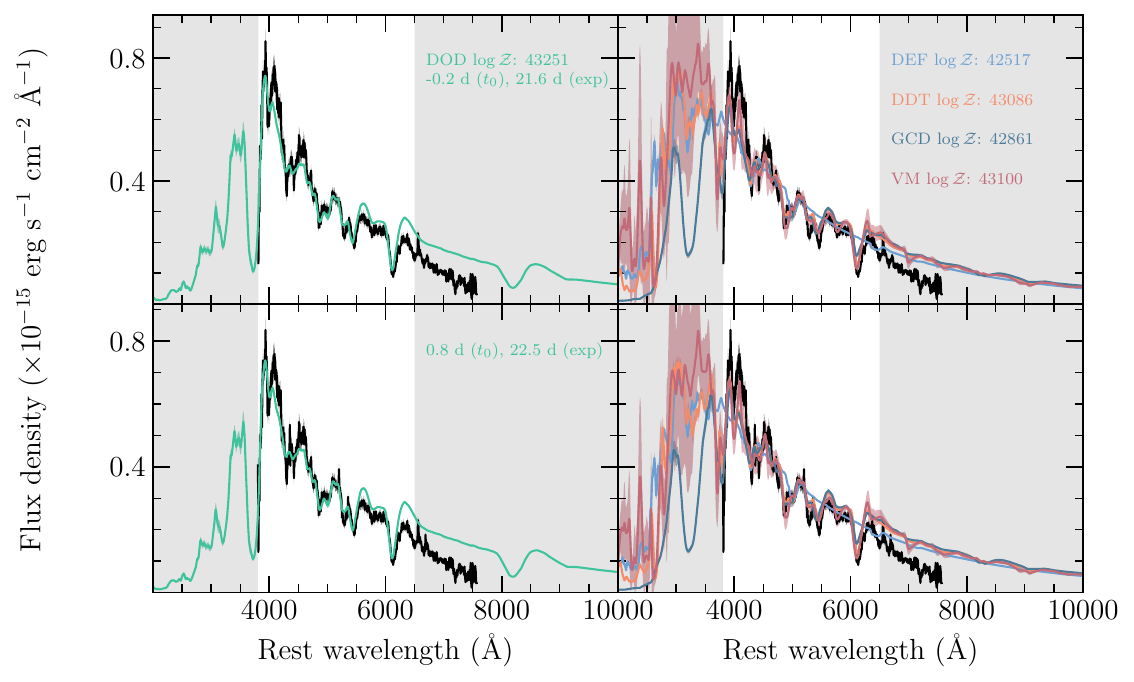}
    \caption{Fits to ZTF19adbnfne as in Fig.~\ref{fig:ZTF18aahheaj_fit}}
    \label{fig:ZTF19adbnfne_fit}
\end{figure}

\begin{figure}
    \centering
    \includegraphics[width=\columnwidth]{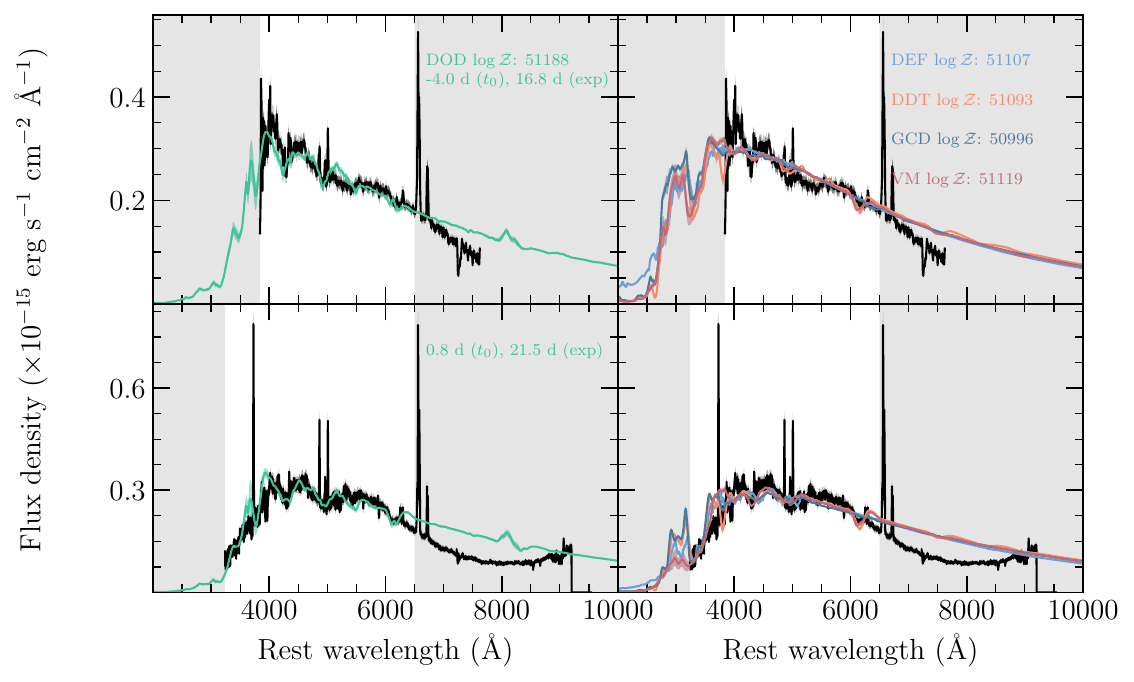}
    \caption{Fits to ZTF20aadfyjl as in Fig.~\ref{fig:ZTF18aahheaj_fit}}
    \label{fig:ZTF20aadfyjl_fit}
\end{figure}

\begin{figure}
    \centering
    \includegraphics[width=\columnwidth]{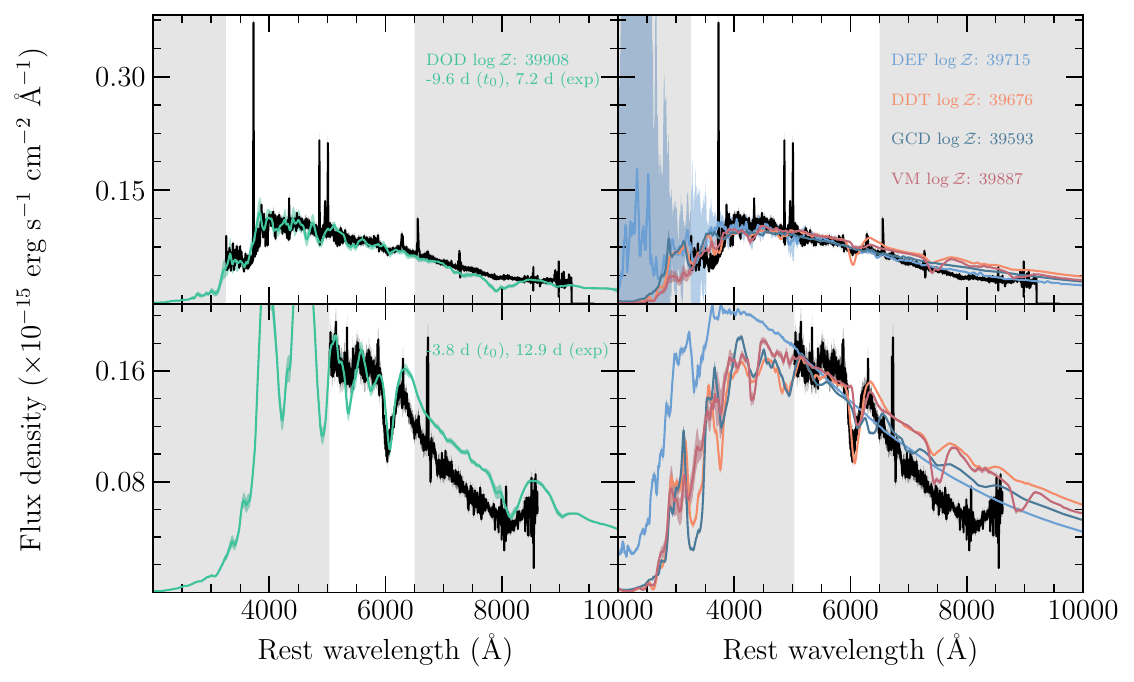}
    \caption{Fits to ZTF20aafsnpp as in Fig.~\ref{fig:ZTF18aahheaj_fit}}
    \label{fig:ZTF20aafsnpp_fit}
\end{figure}

\begin{figure}
    \centering
    \includegraphics[width=\columnwidth]{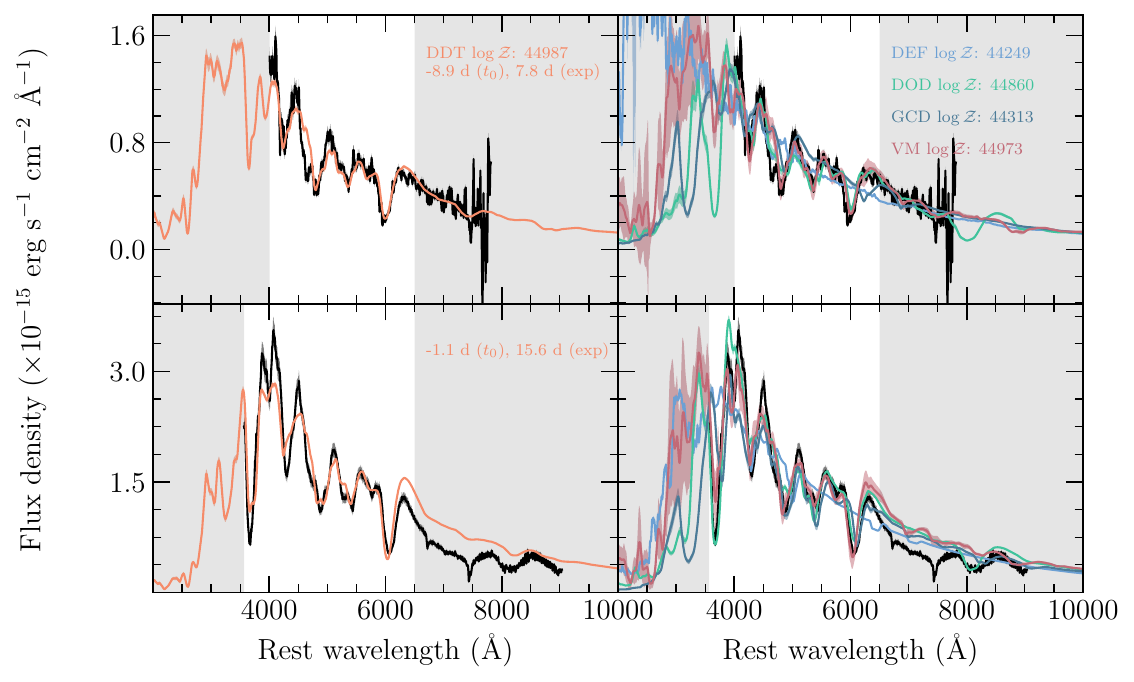}
    \caption{Fits to ZTF20acikuon as in Fig.~\ref{fig:ZTF18aahheaj_fit}}
    \label{fig:ZTF20acikuon_fit}
\end{figure}

\section{Primary sample tables}
\label{sect:primary_tables}
\onecolumn

\begin{center}
\renewcommand*{\arraystretch}{1.5}
\begin{longtable}{lcclccc}
\caption{Fitted properties of ZTF sample}  \\
\hline
\multicolumn{1}{l}{\textbf{Name}}	 &	  \multicolumn{1}{c}{\textbf{Redshift}}	 & \multicolumn{1}{c}{\textbf{No. of}}	 &	\multicolumn{1}{l}{\textbf{Best}}	 &	\multicolumn{1}{c}{\textbf{Explosion epoch}}	 &	\multicolumn{1}{c}{\textbf{Distance modulus}}	 &	\multicolumn{1}{c}{\textbf{Host extinction}}	\\
        &	      &	\multicolumn{1}{c}{\textbf{spectra}}	    &	\multicolumn{1}{l}{\textbf{model}}	 &	\multicolumn{1}{c}{\textbf{(MJD)}}	 &	\multicolumn{1}{c}{\textbf{(mag)}}	 &	\multicolumn{1}{c}{\textbf{$E(B-V)$}}	\\ \hline\hline
\label{tab:fitted_sn_properties}
\endfirsthead
\multicolumn{7}{c}%
{{\tablename\ \thetable{} -- continued from previous page}} \\
\hline
\multicolumn{1}{l}{\textbf{Name}}	 &	  \multicolumn{1}{c}{\textbf{Redshift}}	 & \multicolumn{1}{c}{\textbf{No. of}}	 &	\multicolumn{1}{l}{\textbf{Best}}	 &	\multicolumn{1}{c}{\textbf{Explosion epoch}}	 &	\multicolumn{1}{c}{\textbf{Distance modulus}}	 &	\multicolumn{1}{c}{\textbf{Host extinction}}	\\
        &	      &	\multicolumn{1}{c}{\textbf{spectra}}	    &	\multicolumn{1}{l}{\textbf{model}}	 &	\multicolumn{1}{c}{\textbf{(MJD)}}	 &	\multicolumn{1}{c}{\textbf{(mag)}}	 &	\multicolumn{1}{c}{\textbf{$E(B-V)$}}	\\ \hline\hline
\endhead
\hline \multicolumn{7}{|r|}{{Continued on next page}} \\ \hline
\endfoot
\endlastfoot
ZTF18aagtcxj & 0.032 & 2 & VM & 58188.26~$^{58188.71}_{58187.81}$ & 35.86~$^{35.94}_{35.79}$ & 0.296~$^{0.321}_{0.272}$  \\
ZTF18aahheaj & 0.053 & 3 & VM & 58203.76~$^{58204.40}_{58203.12}$ & 36.99~$^{37.04}_{36.95}$ & 0.000~$^{0.003}_{0.000}$  \\
ZTF18aaumeys & 0.036 & 2 & DDT & 58247.55~$^{58248.05}_{58247.04}$ & 36.43~$^{36.51}_{36.35}$ & 0.002~$^{0.009}_{0.000}$  \\
ZTF18abauprj & 0.024 & 2 & DDT & 58282.30~$^{58282.75}_{58281.85}$ & 35.01~$^{35.07}_{34.95}$ & 0.000~$^{0.001}_{0.000}$  \\
ZTF18abosdwf & 0.056 & 2 & VM & 58331.13~$^{58332.10}_{58330.17}$ & 36.81~$^{36.87}_{36.75}$ & 0.001~$^{0.003}_{0.000}$  \\
ZTF18abuhzfc & 0.038 & 2 & VM & 58360.76~$^{58361.66}_{58359.85}$ & 36.04~$^{36.16}_{35.91}$ & 0.102~$^{0.135}_{0.068}$  \\
ZTF18acbxsge & 0.036 & 3 & GCD/DDT & 58419.65~$^{58420.12}_{58419.18}$ & 35.81~$^{35.89}_{35.72}$ & 0.000~$^{0.002}_{0.000}$  \\
ZTF18acurmpd & 0.029 & 2 & DOD & 58216.81~$^{58217.38}_{58216.25}$ & 35.71~$^{35.81}_{35.61}$ & 0.001~$^{0.004}_{0.000}$  \\
ZTF18adaifep & 0.042 & 2 & DDT & 58469.14~$^{58469.73}_{58468.55}$ & 36.25~$^{36.32}_{36.19}$ & 0.000~$^{0.002}_{0.000}$  \\
ZTF19aabmybj & 0.040 & 2 & DOD & 58480.45~$^{58481.10}_{58479.81}$ & 36.54~$^{36.63}_{36.46}$ & 0.000~$^{0.002}_{0.000}$  \\
ZTF19aabvfwn & 0.041 & 2 & DOD & 58484.01~$^{58484.24}_{58483.78}$ & 36.65~$^{36.72}_{36.57}$ & 0.001~$^{0.006}_{0.000}$  \\
ZTF19aadnxat & 0.032 & 2 & DOD & 58497.79~$^{58497.96}_{58497.62}$ & 35.64~$^{35.66}_{35.61}$ & 0.001~$^{0.003}_{0.000}$  \\
ZTF19aadoocn & 0.041 & 2 & GCD/DDT & 58487.87~$^{58489.06}_{58486.69}$ & 36.56~$^{36.64}_{36.48}$ & 0.003~$^{0.012}_{0.000}$  \\
ZTF19aagkfvq & 0.053 & 2 & DOD & 58964.56~$^{58965.72}_{58963.39}$ & 37.21~$^{37.29}_{37.13}$ & 0.002~$^{0.009}_{0.000}$  \\
ZTF19aakoasy & 0.033 & 2 & DDT & 58530.96~$^{58531.55}_{58530.37}$ & 35.90~$^{35.99}_{35.82}$ & 0.000~$^{0.002}_{0.000}$  \\
ZTF19aakzwao & 0.019 & 3 & DOD & 58539.46~$^{58539.76}_{58539.17}$ & 35.10~$^{35.21}_{35.00}$ & 0.000~$^{0.002}_{0.000}$  \\
ZTF19aamkdtq & 0.034 & 2 & VM & 58537.15~$^{58538.23}_{58536.08}$ & 35.94~$^{36.08}_{35.80}$ & 0.064~$^{0.107}_{0.022}$  \\
ZTF19aampqcq & 0.019 & 2 & GCD & 58550.93~$^{58551.69}_{58550.17}$ & 35.30~$^{35.37}_{35.22}$ & 0.006~$^{0.020}_{0.000}$  \\
ZTF19aamrdmm & 0.042 & 2 & GCD & 58540.52~$^{58541.19}_{58539.85}$ & 36.43~$^{36.51}_{36.35}$ & 0.000~$^{0.002}_{0.000}$  \\
ZTF19aanyuyh & 0.025 & 2 & VM & 59035.89~$^{59036.74}_{59035.04}$ & 34.37~$^{34.42}_{34.32}$ & 0.002~$^{0.008}_{0.000}$  \\
ZTF19aaxffum & 0.055 & 2 & GCD & 58627.06~$^{58627.95}_{58626.17}$ & 37.07~$^{37.19}_{36.95}$ & 0.167~$^{0.186}_{0.149}$  \\
ZTF19acazbqm & 0.032 & 2 & VM & 58743.08~$^{58744.16}_{58742.00}$ & 36.00~$^{36.20}_{35.80}$ & 0.101~$^{0.150}_{0.052}$  \\
ZTF19acxyumq & 0.036 & 2 & DOD & 58824.69~$^{58825.13}_{58824.25}$ & 36.04~$^{36.11}_{35.96}$ & 0.000~$^{0.001}_{0.000}$  \\
ZTF19acyiqnw & 0.033 & 2 & DDT & 58823.80~$^{58824.46}_{58823.14}$ & 36.04~$^{36.12}_{35.95}$ & 0.000~$^{0.002}_{0.000}$  \\
ZTF19adbnfne & 0.056 & 2 & DOD & 58837.21~$^{58838.27}_{58836.15}$ & 37.12~$^{37.18}_{37.06}$ & 0.000~$^{0.001}_{0.000}$  \\
ZTF20aadfyjl & 0.049 & 2 & DOD & 58843.40~$^{58843.99}_{58842.80}$ & 36.50~$^{36.58}_{36.42}$ & 0.240~$^{0.264}_{0.217}$  \\
ZTF20aafsnpp & 0.044 & 2 & DOD & 58858.49~$^{58858.84}_{58858.14}$ & 36.93~$^{37.03}_{36.83}$ & 0.173~$^{0.202}_{0.143}$  \\
ZTF20acikuon & 0.023 & 2 & DDT & 59131.06~$^{59131.65}_{59130.47}$ & 35.01~$^{35.11}_{34.92}$ & 0.001~$^{0.005}_{0.000}$  \\
\hline\hline
\end{longtable}
\end{center}

\clearpage
\twocolumn
\section{Additional figures}
\label{sect:additional_figures}

\begin{figure}
    \centering
    \includegraphics[width=\columnwidth]{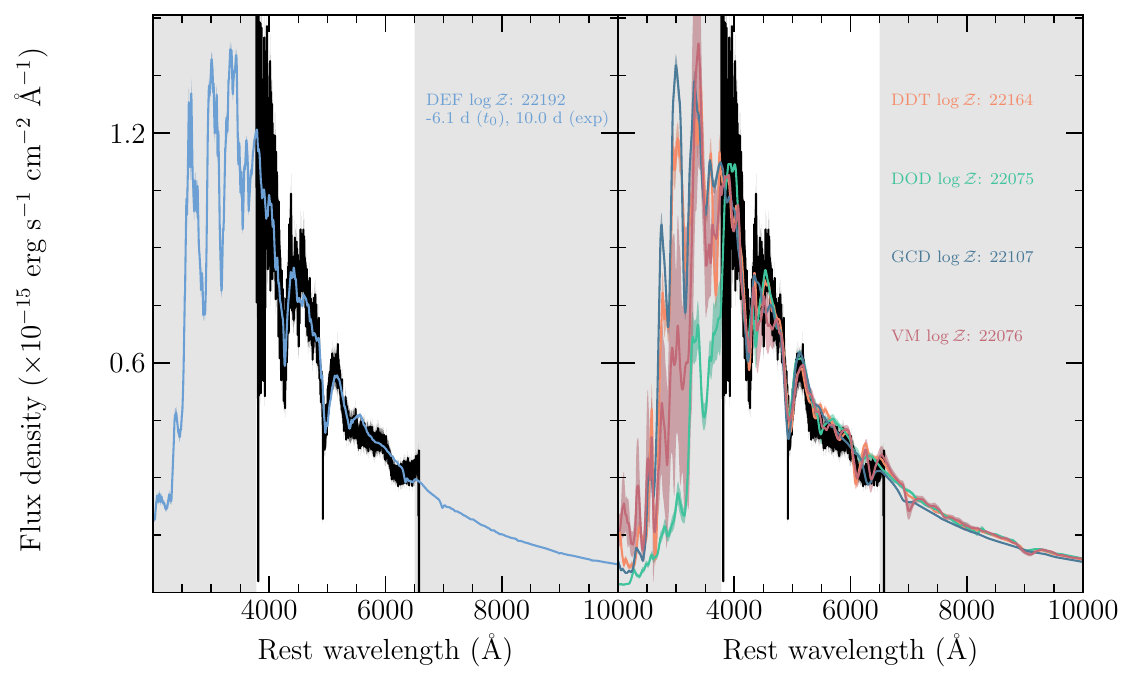}
    \caption{Fits to ZTF19aafmfxg as in Fig.~\ref{fig:ZTF18aahheaj_fit}}
    \label{fig:ZTF19aafmfxg_fit}
\end{figure}

\begin{figure}
    \centering
    \includegraphics[width=\columnwidth]{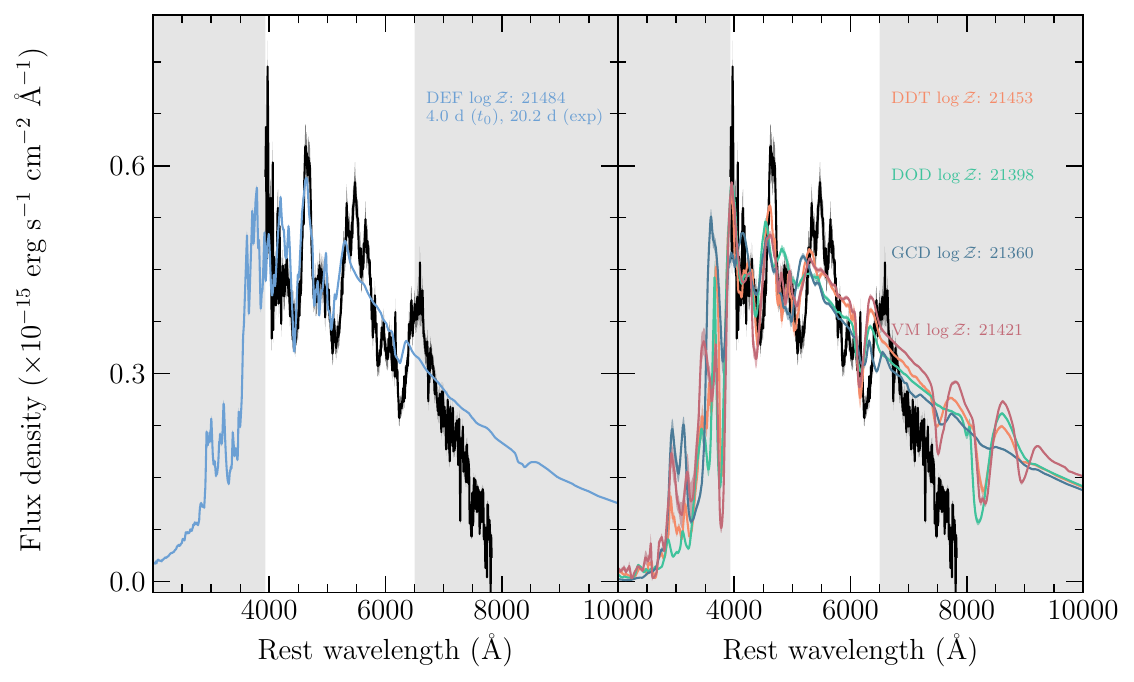}
    \caption{Fits to ZTF19acxgxcu as in Fig.~\ref{fig:ZTF18aahheaj_fit}}
    \label{fig:ZTF19acxgxcu_fit}
\end{figure}


\bsp	
\label{lastpage}
\end{document}